\documentclass[fleqn,usenatbib]{mnras}

\usepackage{newtxtext,newtxmath}

\usepackage[T1]{fontenc}

\DeclareRobustCommand{\VAN}[3]{#2}
\let\VANthebibliography\thebibliography
\def\thebibliography{\DeclareRobustCommand{\VAN}[3]{##3}\VANthebibliography}

\usepackage{graphicx}	
\usepackage{amsmath}	
\defcitealias{Liu2025}{Paper I}

\title[Effectively optically thin flow]{The effectively optically thin accretion flow and its implication in supermassive black holes}

\author[M. Liu, B. Liu, Y. Wang, H. Cheng, and W. Yuan]{
Mingjun Liu,$^{1,2}$
B. F. Liu,$^{1,2}$\thanks{E-mail: bfliu@nao.cas.cn}
Yilong Wang,$^{1,2}$
Huaqing Cheng,$^{1}$
and Weimin Yuan$^{1,2}$
\\
$^{1}$National Astronomical Observatories, Chinese Academy of Sciences, 20A Datun Road, Beijing 100101, People's Republic of China\\
$^{2}$School of Astronomy and Space Science, University of Chinese Academy of Sciences, 19A Yuquan Road, Beijing 100049, People's Republic of China
}

\date{Accepted XXX. Received YYY; in original form ZZZ}

\pubyear{\the\year{}}

\begin{document}
\label{firstpage}
\pagerange{\pageref{firstpage}--\pageref{lastpage}}
\maketitle

\begin{abstract}
Based on a unified description of various accretion flows, we find a long-ignored solution - the effectively optically thin accretion flow, occurring at accretion rates around Eddington value. As a consequence of radiation-pressure dominance,  the density in a standard thin disc (SSD) decreases with the increase of accretion rates, making the innermost region effectively optically thin. Further increase in accretion rate leads to a rise of the temperature so that the Compton cooling is able to balance the accretion released energy. We demonstrate that the effectively optically thin flow is characterized by moderate temperature and large scattering optical depth, producing a multi-color Wien spectrum. For an appropriate accretion rate, the accretion flow transforms from an outer SSD to an inner effectively optically thin flow. Thus, the spectra of the whole accretion flow exhibit two components, i.e., a multi-color Wien spectrum at higher frequency and a multi-color blackbody, the former could provide an alternative origin of soft X-ray excess or formation of warm corona in active galactic nuclei (AGNs). Our stability analysis proves it is thermally stable and viscously unstable, indicating its existence in accreting systems. We show that effectively optically thin accretion flow exists in supermassive black holes for accretion rates around 0.1 to 10 times Eddington value, bridging the SSD at low accretion rates and slim disc at high rates. By comparing the predictions and average spectra of AGN, we constrain the viscosity parameter to be $\alpha \sim 0.03$, in good agreement with that derived from  observed variability.
\end{abstract}




\section{Introduction}\label{sec:intro}
The standard thin disc \citep[SSD, e.g.,][]{Shakura_Sunyaev_1973} and slim disc \citep[e.g.,][]{Katz_1977,Abramowicz_1988} describe the cold accretion flow around black holes 
at sub- and super-critical accretion phases, respectively. Both are generally expected to be optically thick, interpreting the multicolor blackbody component in spectra of various accreting systems, such as the high/soft state of X-ray binaries \citep[XRBs, e.g.][and references therein]{Remillard_McClintock_2006}, the big blue bump in active galactic nuclei \citep[AGNs, e.g.,][]{Malkan_Sargent_1982,Malkan_1983,Elvis_1994}, the soft X-ray emission of tidal disruption events \citep[TDEs, e.g.,][]{Rees_1988,Wen_2020} and the ultraluminous X-ray sources \citep[ULXs, e.g.,][]{Kaaret_2017}.

In contrast to the extensive studies and wide application of the SSD and slim disc, another solution that exists at high accretion rates has been nearly forgotten, that is, the effectively optically thin accretion flows. This solution was mentioned in the pioneering work of SSD by \citet{Shakura_Sunyaev_1973} and \citet{Novikov_Thorne_1973}. After that, a few efforts were made to apply this effectively optically thin accretion flow to explain the hard X-ray emission of Cygnus X-1 \citep{Thorne_Price_1975,Lightman_Shapiro_1975} and the UV/soft X-ray spectra of AGNs \citep{Callahan_1977,Czerny_Elvis_1987,Wandel_Petrosian_1988}. However, the effectively optically thin accretion flow, characterized by the low optical depth and radiation pressure support, has long been speculated to be thermally unstable \citep[e.g.,][]{Callahan_1977,Sakimoto1981}. Furthermore, these studies on the effectively optically thin accretion flow suffered the complexity of radiative transfer in accretion flows. Consequently, this topic soon faded into obscurity.

After a decade, the importance of the effectively optically thin accretion flows in super-Eddington accretion flows was realized in several works. \citet{Beloborodov_1998} encountered the effectively optically thin accretion flow and quantitatively described its structure for the first time. Following that, this accretion flow was investigated by \citet{Chen_Wang_2004,Artemova_2006,Soria_2008,Klepnev_Bisnovatyi-Kogan_2010}. Since then no further studies and applications have been found.

Theoretically, the effectively optically thin accretion flow does occur at accretion rates around the Eddington rate, connecting the radiation-pressure-dominant SSD at lower accretion rates and the slim disc at higher accretion rates in the solution regime sorted by the accretion rate. With the increase of the accretion rates, the density in a radiation-pressure-dominant SSD decreases, thereby suppressing the bremsstrahlung and its self-absorption. At a certain critical accretion rate, the accretion-liberated energy can no longer be radiated and then heats the gas until the temperature is sufficiently high so that Comptonization plays a role in cooling. Consequently, the accretion flow becomes effectively optically thin with a temperature higher than that in the SSD. 

Nevertheless, the regime for the existence of the effectively optically thin accretion flow has not been well determined yet. It was predicted to occur when the effective optical depth is less than one derived from the solutions of either the SSD or the slim disc.  

In a recent work \citep{Liu2025}, hereafter \citetalias{Liu2025}, we developed a generalized description of accretion flows, unifying all the possible solutions. This makes it possible to investigate the effectively optically thin accretion flows self-consistently, in particular, to determine the regime of parameters for its existence. The comparison with observations would examine whether it occurs true in practice and constrains the model parameters. 

In this work, we revisit the effectively optically thin accretion flow on the basis of the generalized solution \citetalias{Liu2025}. We present a self-consistent description, perform stability analysis, and calculate the spectra for the first time. The model is briefly illustrated in Section \ref{sec:model}. The properties of the effectively optically thin accretion flow are presented in Section \ref{sec:results} and discussed in Section \ref{sec:discussion}. The main conclusions are made in Section ~\ref{sec:conclusion}.
 
\section{Model description and the basic equations} \label{sec:model}

We consider the steady axisymmetric accretion flow around a black hole under the Newtonian framework. We generalize the equations to describe all possible accretion processes, where the entropy advection, the radiation pressure, and the photon trapping effect are all included self-consistently. As has been shown in the previous study \citepalias[][]{Liu2025}, with the generalized equations the advection-dominated accretion flow \citep[ADAF, e.g.,][]{Ichimaru_1977,Narayan_Yi_1994,Narayan_Yi_1995}, the Shapiro-Lightman-Eardley solution \citep[SLE,][]{Shapiro_1976}, SSD, and slim disc branches can be well reproduced in a wide range of accretion rates from sub- to super-Eddington accretion. In addition to unifying these solutions, the model consistently describes the transition from SSD to slim disc with an increase of accretion rate, and the transition from an SSD to a slim disc from outer to inner regions at fixed accretion rates. The accretion flow is described by the following generalized equations, i.e., the continuity equation (\ref{eq:continuity}), radial momentum equation (\ref{eq:momentum}), azimuthal momentum equation (\ref{eq:angular}), vertical hydrostatic equilibrium equation (\ref{eq:vertical}), total energy equation (\ref{eq:energy_total}), and energy equation for electrons (\ref{eq:energy_electron}) \citepalias[for details see][]{Liu2025}
\begin{equation}
    \dot{M}=-4\pi \rho RH\upsilon,
    \label{eq:continuity}
\end{equation}
\begin{equation}
    \upsilon\frac{\text d\upsilon}{\text dR}-\Omega^2R=-\Omega^2_{\text K}R-\frac{1}{\rho}\frac{\text dp}{\text dR},
    \label{eq:momentum}
\end{equation}
\begin{equation}
    \rho RH\upsilon\frac{\text d \left(\Omega R^2\right)}{\text dR}=\frac{\text d}{\text dR}\left(\nu\rho R^3H\frac{\text d\Omega}{\text dR}\right),
    \label{eq:angular}
\end{equation}
\begin{equation}
    \frac{p}{\rho H}=\frac{2}{5}\Omega_\mathrm{K}^2H,
    \label{eq:vertical}
\end{equation}
\begin{equation}
    \rho\upsilon\frac{\text de_\mathrm{int}}{\text dR}-\upsilon\frac{p}{\rho}\frac{\text d\rho}{\text dR}=\nu\rho \left(R\frac{\text d\Omega}{\text dR}\right)^2-q_\mathrm{rad}.
     \label{eq:energy_total}
\end{equation}
\begin{equation}
    q^\mathrm{ie}=q_\mathrm{rad},
    \label{eq:energy_electron}
\end{equation}
where $G$ is the gravitational constant, $\dot{M}$ the accretion rate, $R$ the distance, $M$ the mass of the accreting object, $\rho$ the density, $H$ the vertical scale height, $\Sigma\equiv2\rho H$ the surface density, $p=p_\mathrm{g}+p_\mathrm{m}+p_\mathrm{r}$ the total pressure composed of the gas pressure, magnetic pressure and radiation pressure, $T_\mathrm{i}$ and $T_\mathrm{e}$ the temperatures, $\upsilon$ the radial velocity, $\Omega$ the angular velocity, $\Omega_\mathrm{K}\equiv(GM/R^3)^{1/2}$ the Keplerian angular velocity, $c_\mathrm{s}\equiv(p/\rho)^{1/2}$ the effective isothermal sound speed, $\tau=\tau_\mathrm{es}+\tau_\mathrm{abs}$ the total optical depth, $\tau_\mathrm{eff}\equiv\left[\tau_{\rm abs}(\tau_\mathrm{abs}+\tau_{\rm es})\right]^{1/2}$ the effective optical depth, $\nu=\alpha c_\mathrm{s}^2/\Omega_\mathrm{K}$ the kinematic viscosity, $\alpha$ the viscosity parameter. The internal energy $e_\mathrm{int}$ and its relation with the pressure, the energy transfer rate between ions and electrons $q^\mathrm{ie}$, and the height-averaged net radiation rate and its relation with flux $F_\mathrm{rad}$, $q_\mathrm{rad}\equiv F_\mathrm{rad}/H$, are expressed as follows,

\begin{equation}\label{eq:int}
    e_\mathrm{int}=\frac{1}{\gamma-1}\frac{p_\mathrm{g}}{\rho }+\frac{3p_\mathrm{m}}{\rho}+\frac{3p_\mathrm{r}}{\rho}\equiv\frac{1}{\Gamma_3-1}\frac{p}{\rho},
\end{equation}
\begin{equation}
    p_{\rm g}=n_\mathrm{i}k_\mathrm{B}T_\mathrm{i}+n_\mathrm{e}k_\mathrm{B}T_\mathrm{e},\, p_{\rm m}=\frac{B^2}{24\pi}, \, p_{\rm r}=\frac{F_\mathrm{rad}}{2c}\left(\tau+\frac{2}{\sqrt{3}}\right),
    \label{eq:EoS}
\end{equation}
\begin{equation}\label{eq:qie}
    \begin{aligned}
        q^{\mathrm{ie}}&=\frac{3m_\mathrm{e}}{2m_\mathrm{p}}\sum_j Z^2_jn_jn_\mathrm{e} \sigma_\mathrm{T}c\frac{k_\mathrm{B}(T_\mathrm{i}-T_\mathrm{e})}{K_2(1/\theta_\mathrm{i})K_2(1/\theta_\mathrm{e})}\ln{\Lambda}\\
        &\times\left[\frac{2(\theta_\mathrm{i}+\theta_\mathrm{e})^2+1}{\theta_\mathrm{i}+\theta_\mathrm{e}}K_1\left(\frac{\theta_\mathrm{i}+\theta_\mathrm{e}}{\theta_\mathrm{i}\theta_\mathrm{e}}\right)+2K_0\left(\frac{\theta_\mathrm{i}+\theta_\mathrm{e}}{\theta_\mathrm{i}\theta_\mathrm{e}}\right)\right],
    \end{aligned}
\end{equation}
\begin{equation}
    F_\mathrm{rad}=4\sigma T_\mathrm{e}^4\left(\frac{3}{2}\tau+\sqrt{3}+\frac{1}{\tau_\mathrm{abs}}\right)^{-1},
    \label{eq:q_rad}
\end{equation}
\begin{equation}
    \tau_\mathrm{abs}=\frac{\left(q_\mathrm{br}+q_\mathrm{syn}+q_\mathrm{br,C}+q_\mathrm{syn,C}\right)H}{4\sigma T_\mathrm{e}^4},
    \label{eq:abs}
\end{equation}
where $\gamma=5/3$ is the ratio of specific heats for ideal gas, $\Gamma_3$ is the general adiabatic exponent by inclusion of the radiation pressure, magnetic pressure and gas pressure \citep[e.g.,][]{Chandrasekhar_1967,Clayton_1983,Mihalas_1984}, the magnetic field $B$ is parameterized by $\beta$ assuming magnetic pressure in partition with the gas pressure, $\beta\equiv p_\mathrm{g}/(p_\mathrm{g}+p_\mathrm{m})$, $k_\mathrm{B}$ is the Boltzmann constant, $c$ is the speed of light, $\sigma_\mathrm{T}$ is the Thomson scattering cross-section, $n_\mathrm{i}$ and $n_\mathrm{e}$ are the number densities, $m_\mathrm{p}$ and $m_\mathrm{e}$ are the mass of proton and electron, respectively, $Z_j$ is the charge number of the $j$-th species, $n_j$ is the number density of the $j$-th species, i.e., $\sum_j n_j=n_\mathrm{i}$, the Coulomb logarithm $\ln\Lambda$ is taken as 20 in this work, $\theta_\mathrm{i}=k_\mathrm{B}T_\mathrm{i}/m_\mathrm{i}c^2$ and $\theta_\mathrm{e}=k_\mathrm{B}T_\mathrm{e}/m_\mathrm{e}c^2$ are the dimensionless ion and electron temperatures, respectively, $K_n$ is the $n$-th order modified Bessel function, $\sigma$ is the Stefan-Boltzmann constant, $q_\mathrm{br}$, $q_\mathrm{syn}$, $q_\mathrm{br,C}$, and $q_\mathrm{syn,C}$ are the emissivity of bremsstrahlung, synchrotron, and self-Compton scattering  \citep[for details see][]{Liu2025}, from which the absorption optical depth is obtained based on the Kirchhoff's law \citep{Narayan_Yi_1995}.
 
In this work, we improve the treatments of Comptonization $q_\mathrm{br,C}$ and $q_\mathrm{syn,C}$ so that they can be better described for both cases of the effectively optically thin and thick. The radiative cooling rates of self-Compton processes are written as follows,
\begin{equation}\label{eq:comp}
    \begin{split}
        &q_\mathrm{br,C}={\frac{q_\mathrm{br}}{{\theta_\mathrm{e}}}}\int_{x_\mathrm{c}}^{\theta_\mathrm{e}}\left(\eta-1\right) \text dx,\\
        &q_\mathrm{syn,C}=
         \begin{cases}
         \left(\eta-1\right)q_\mathrm{syn}, & \nu_\mathrm{br}\leq\nu_\mathrm{syn},\\
         0, & \nu_\mathrm{br}>\nu_\mathrm{syn},
         \end{cases}
    \end{split}
\end{equation}
where $\theta_\mathrm{e}=k_\mathrm{B}T_\mathrm{e}/m_\mathrm{e}c^2$, $x=h\nu/m_\mathrm{e}c^2$, $h$ is the Planck constant, $\nu_\mathrm{c}$ is the self-absorption frequency, $x_\mathrm{c}\equiv h\nu_\mathrm{c}/m_\mathrm{e}c^2$, $\nu_\mathrm{br}$ and $\nu_\mathrm{syn}$ are the critical frequencies of bremsstrahlung and synchrotron, respectively, the average luminosity enhancement factor \citep{Dermer_1991},
\begin{equation}
    \eta=1+\frac{P(A-1)}{1-PA}\left[1-\left(\frac{x}{3\theta_\mathrm{e}}\right)^{-1-\ln P/\ln A}\right],
    \label{eq:eta}
\end{equation}
$P$ is the scattering probability of photons and $A=1+4\theta_\mathrm{e}+16\theta_\mathrm{e}^2$ is the mean amplification factor of photons in single scattering. 

Two improvements are presented in this work, i.e., the self-absorption frequency of soft photons for Comptonization and the scattering probability. Since we mainly treat the accretion flow with weak synchrotron, the self-absorption of bremsstrahlung needs to be included, i.e., the absorption frequency  $\nu_\mathrm{c}=\max\{\nu_\mathrm{syn},\nu_\mathrm{br}\}$, The expression of $\nu_\mathrm{syn}$ follows \citepalias[][]{Liu2025}.  The critical frequency $\nu_\mathrm{br}$ is chosen as the frequency $\nu_\mathrm{ff}$ where the bremsstrahlung becomes optically thick, i.e., $\tau_\mathrm{abs}^\mathrm{ff}=\alpha_\nu^\mathrm{ff}H=1$. The free-free absorption coefficient $\alpha_\nu^\mathrm{ff}$ is computed following \citet{Rybicki_Lightman_1979},
\begin{equation}
    \begin{split}
        \alpha_\nu^\mathrm{ff}=\frac{4e^6}{3m_\mathrm{e}hc}\left(\frac{2\pi}{3k_\mathrm{B}m_\mathrm{e}}\right)^{1/2}\sum_jZ^2_jn_jn_\mathrm{e}T_\mathrm{e}^{-1/2}\\
        \times\frac{1-\exp{\left(-h\nu/kT_\mathrm{e}\right)}}{\nu^{3}}\bar{g}_\mathrm{ff},
    \end{split}
\end{equation}
where $e$ is the electron charge and the Gaunt factor $\bar{g}_\mathrm{ff}=\sqrt{3}\pi^{-1}\exp{(x/2\theta_\mathrm{e})}K_0(x/2\theta_\mathrm{e})$ \citep{Greene_1959,Brussaard_1962,Mewe_1986}.

The treatment on scattering given by \citet{Dermer_1991} is only suited for low scattering optical depth and saturated scattering. While \citet{Sunyaev_Titarchuk_1980,Liang_Nolan_1984} present the reasonable expression of scattering. Therefore, we combine these work to generalize the expression of scattering probability, i.e., $P=\min{\left[1-\exp\left(-\tau_\mathrm{es,eff}\right),\exp\left(-\beta_\mathrm{C}\right)\right]}$, where $\beta_\mathrm{C}=\pi^2/12\left(\tau_\mathrm{es, eff}+2/3\right)^2$ and $\tau_\mathrm{es,eff}$ is the effective scattering optical depth. $\tau_\mathrm{es,eff}$ is evaluated via the effective mean free path of photons \citep{Rybicki_Lightman_1979}, i.e., $\tau_\mathrm{es,eff}=\tau_\mathrm{es}\min\left[1,1/\tau_\mathrm{eff,r}\right]$ with $\tau_\mathrm{eff,r}=[\tau_\mathrm{abs,r}(\tau_\mathrm{es}+\tau_\mathrm{abs,r})]^{1/2}$. The absorption $\tau_\mathrm{abs,r}$ only includes the effects of bremsstrahlung and synchrotron, i.e., $\tau_\mathrm{abs,r}=\left(q_\mathrm{br}+q_\mathrm{syn}\right)H/4\sigma T_\mathrm{e}^4$. 

We obtain temperatures $T_\mathrm{i}$, $T_\mathrm{e}$, advection fraction $f$, and general adiabatic exponent $\Gamma_3$ through solving the two energy equations for the total accretion flows (Eq.\ref{eq:energy_total}) and the electrons alone (Eq.\ref{eq:energy_electron}), the definition equation of $\Gamma_3$ (Eq.\ref{eq:int}), and the state of equation (Eq.\ref{eq:EoS}) according to the numerical method in Section 2.3 of \citetalias{Liu2025}. Given $T_\mathrm{i}$, $T_\mathrm{e}$, $f$, and $\Gamma_3$, other physical quantities can be derived from the equations in this Section. 

The spectra in this work are estimated theoretically since we mainly deal with cases with large scattering optical depths for which the Monte Carlo method is unfeasible. Furthermore, we need to treat the spectra with Comptonization in both optically thin and thick situations. Fortunately, \citet{Czerny_Elvis_1987} investigated the spectra of accretion disc with Comptonization. \citet{Manmoto_1997} provided a framework for calculating the spectra of optically thin flow. Following them, we separate the radiative transfer into two steps, i.e., calculating the emergent spectra without Comptonization and then estimating the Comptonized spectra. First, the vertical radiative transfer process which includes emission, absorption and Thomson scattering produces the following emergent spectra, i.e., the Eq.(26) of \citet{Manmoto_1997},
\begin{equation}
    F_{\rm{uncomp,}\nu}=\frac{2\pi}{\sqrt{3}}B_\nu\left[1-\exp{\left(-2\sqrt{3}\tau_{\rm{eff,}\nu}\right)}\right],
\end{equation}
where $B_\nu$ is the Planck function, $\tau_{\rm{eff,}\nu}=[\tau_{\rm{abs,}\nu}(\tau_\mathrm{es}+\tau_{\rm{abs,}\nu})]^{1/2}$, $\tau_{\rm{abs,}\nu}=(q_{\text{br,}\nu}+q_{\text{syn,}\nu})H/4\pi B_\nu$ and the cooling rates $q_{\text{br,}\nu}$, $q_{\text{syn,}\nu}$ are expressed following \citet{Manmoto_1997} at the color temperature $T_\mathrm{col}$. The color temperature is defined as the temperature at the height $H_\mathrm{col}$ under the surface of one effective mean free path. In the effectively optically thin case, this height is taken as the mid-plane. We calculate the color temperature through the vertical structure given by \citet{Hubeny_1990},
\begin{equation}
    T_\mathrm{col}^4=\frac{1}{4}T_\mathrm{eff}^4\left[3\tau_\mathrm{col}\left(1-\frac{\tau_\mathrm{col}}{2\tau}\right)+\sqrt{3}+\frac{1}{\tau_\mathrm{abs}}\right],
\end{equation}
where the optical depth from $H_\mathrm{col}$ to surface is $\tau_\mathrm{col}=\tau\min[1,1/\tau_\mathrm{eff}]$ and the effective temperature $F_\mathrm{rad}=\sigma T_\mathrm{eff}^4$.

Then, we consider the Comptonization of photons following \citet{Czerny_Elvis_1987}. Among all the soft photons for Comptonization, a fraction $f_{\text{sat,}\nu}$ of these photons experience the saturated Compton scattering to produce a Wien bump $W_\nu$. The remaining part of photons which suffers from unsaturated scattering cannot modify the flat spectrum of bremsstrahlung and can only uniformly change the intensity in each band. Thus, the over-all emergent spectrum can be expressed as,
\begin{equation}
    F_\nu=CF_{\rm{uncomp,}\nu}\left(1-f_{\text{sat,}\nu}\right)+W_\nu,
    \label{eq:flux}
\end{equation}
where the normalization $C$ is induced for the effect of unsaturated scattering and the energy conservation. Following \citet{Dermer_1991}, the saturated fraction of photons can be derived as,
\begin{equation}
    f_{\text{sat,}\nu}=\left(\frac{x}{3\theta_\mathrm{col}}\right)^{-\ln{P}/\ln{A}}.
\end{equation}
The photons after the saturated Compton scattering obey the Bose-Einstein (B-E) distribution at the color temperature and the fugacity $z_\gamma$,
\begin{equation}
    W_\nu=\frac{2\pi h\nu^3}{c^2}\frac{1}{z_\gamma^{-1}e^{h\nu/k_\mathrm{B}T_\mathrm{col}}-1}.
    \label{eq:wien}
\end{equation}
When the absorption dominates over the scattering, the Wien spectrum $W_\nu$ degenerates to the blackbody radiation $B_\nu$, i.e., $z_\gamma=1$. The normalization $C$ and fugacity $z_\gamma$ can be derived from the energy conservation through the effective temperature and the Bose integral \citep{Blundell_Blundell_2009},
\begin{equation}
    \begin{split}
        &\int_0^\infty{W_\nu\text d\nu}=\int_0^\infty{CF_{\rm{uncomp,}\nu}f_{\text{sat,}\nu}\frac{3k_\mathrm{B}T_\mathrm{col}}{h\nu}\text d\nu},\\
        &\int^\infty_0{F_\nu\text d\nu}=\sigma T_\mathrm{eff}^4,\\
        &\int_0^\infty{W_\nu\text d\nu}=\frac{\text{Li}_4(z)}{\zeta(4)}\sigma T_\mathrm{col}^4,
    \end{split}
\end{equation}
with polylogarithm function (de Jonquière's function) $\text{Li}_4(z)$ and Riemann zeta function $\zeta(4)=\pi^4/90$.

\section{Numerical results} \label{sec:results}

It has been demonstrated in \citetalias{Liu2025} that there are four patterns of accretion flows around a stellar-mass black hole, that is, the ADAF, SLE, SSD, and slim disc. This work focuses on the potential new pattern, the effectively optically thin accretion flows.  As this solution occurs only when the pressure is dominated by radiation pressure, we choose the supermassive black hole as our main studying object and examine the accretion solution in the innermost region where radiation pressure is often dominant at accretion rates around the Eddington rate. In the following the mass of central black hole, the accretion rate and radius are scaled by solar mass, Eddington rate, and Schwarzschild radius respectively, i.e., $m=M/M_\odot$, $\dot m=\dot M/\dot M_{\rm Edd}$, $r=R/R_{\rm S}$ with $M_\odot=1.989\times 10^{33}\ {\rm g}$, $\dot{M}_\mathrm{Edd}=1.4\times 10^{18}(M/M_\odot) \, {\rm g/s}$ and $R_\mathrm{S}=2.95\times 10^5 M/M_\odot\, {\rm cm}$.

\subsection{General properties of the effectively optically thin flow} \label{subsec:mdot}

\begin{figure*}
\centering 
    \includegraphics[width=\textwidth]{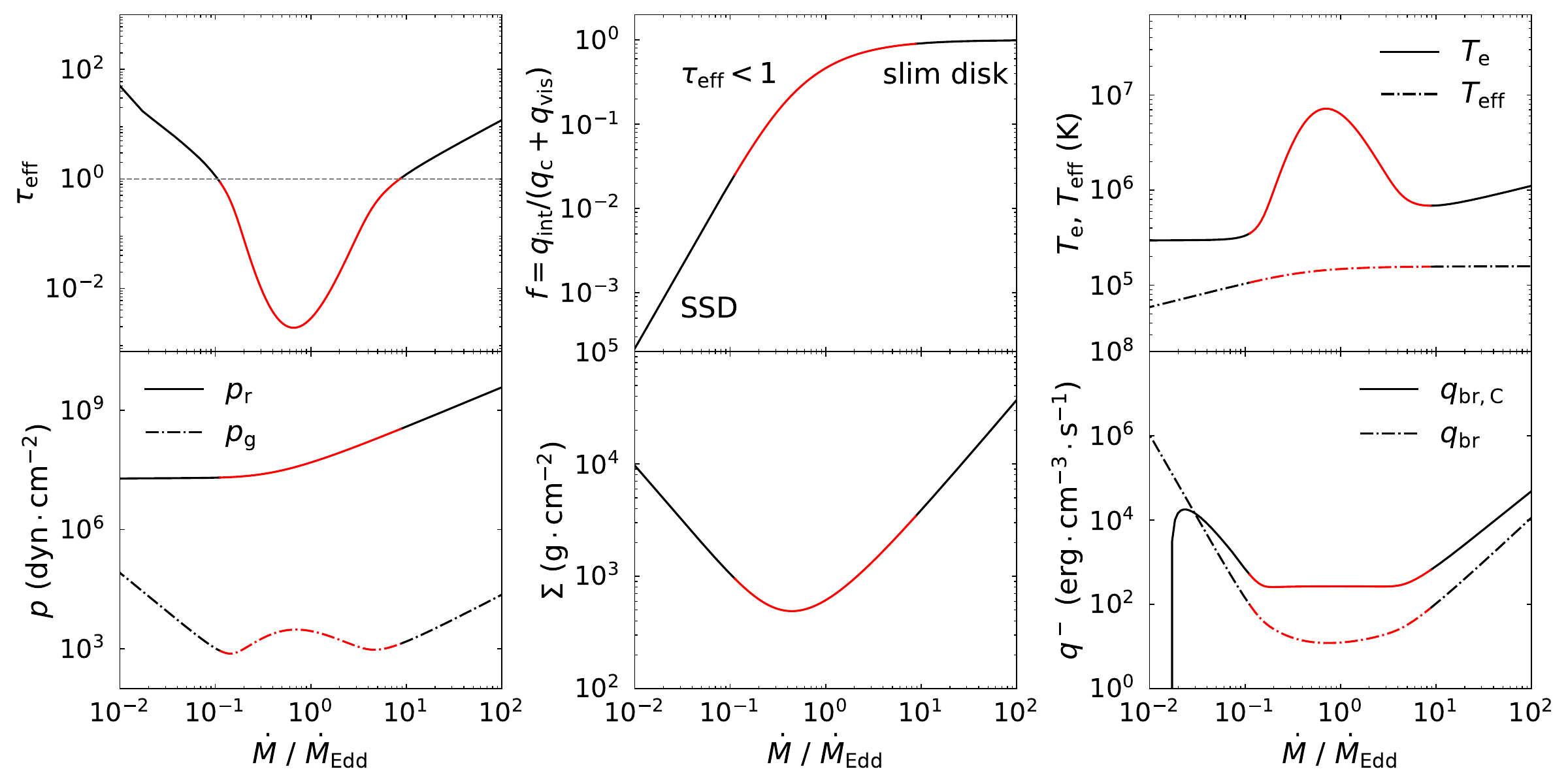}
    \caption{The effective optical depth, advection fraction of energy, electron temperatures at mid-plane (mid-plane temperature hereafter), pressures, surface density, and volume cooling rates of the solution with effectively optically thin accretion flow as the functions of mass accretion rates at the distance of 5 $R_\mathrm{S}$ for $m=10^8$, $\alpha=0.1$, and $p_\mathrm{m}=p_\mathrm{g}$. The effectively optically thin parts are marked in red.}
    \label{fig:regime}
\end{figure*}

For a 10$^8\ M_\odot$ black hole, standard viscosity parameter $\alpha=0.1$ and equipartition magnetic field $\beta=0.5$, we plot in Figure \ref{fig:regime} the structure of the accretion flows at the inner region of $r=5$ for accretion rate ranging from $10^{-2}$ to $10^{2}$ Eddington value. As shown in the upper left panel, the effective optical depth is indeed less than 1 for the accretion rate in a range of $\dot m \sim 0.1-10$. The corresponding structural variables are marked in red in all the panels of Figure \ref{fig:regime}. The new solution setting in between the SSD and slim disc is clearly shown in the upper-middle panel, indicated by the increasing advection fraction towards high accretion rates. The left panels in this figure show the properties of temperature, density, and radiation in the effectively optically thin accretion flow, compared with that of SSD and slim disc. Just as we analyze in the introduction, when the surface density decreases with increasing accretion rate in a radiation pressure dominant SSD (see the left part of the curve in the lower middle panel), the accretion flow eventually becomes effectively optically thin, the gas is heated up to a high temperature (see upper right panel), and Comptonization radiation takes over the bremsstrahlung in cooling (shown in the lower right panel). This trend continues and then turns over, and finally, the accretion flow transits to the slim disc when the accretion rate is sufficiently high so that the accretion flow becomes optically thick ($\tau_{\rm eff}>1$).
 
The most important characteristic of the effectively optically thin accretion flow is the high temperature, which can be as high as $10^7$K around a supermassive black hole for $\alpha=0.1$, two orders of magnitude higher than that in the SSD. It can even reach $\sim 10^9$K for a large viscosity, for example, $\alpha=0.3$. Note that the surface density is much larger than that of a corona or an ADAF, and hence strong X-ray radiation is expected from the effectively optically thin accretion flows. The X-ray luminosity can be comparable to or even higher than the SSD component for a supermassive black hole under some circumstances. For this purpose, we investigate the radial structure of the accretion flows at given accretion rates.

\begin{figure}
\centering 
    \includegraphics[width=\columnwidth]{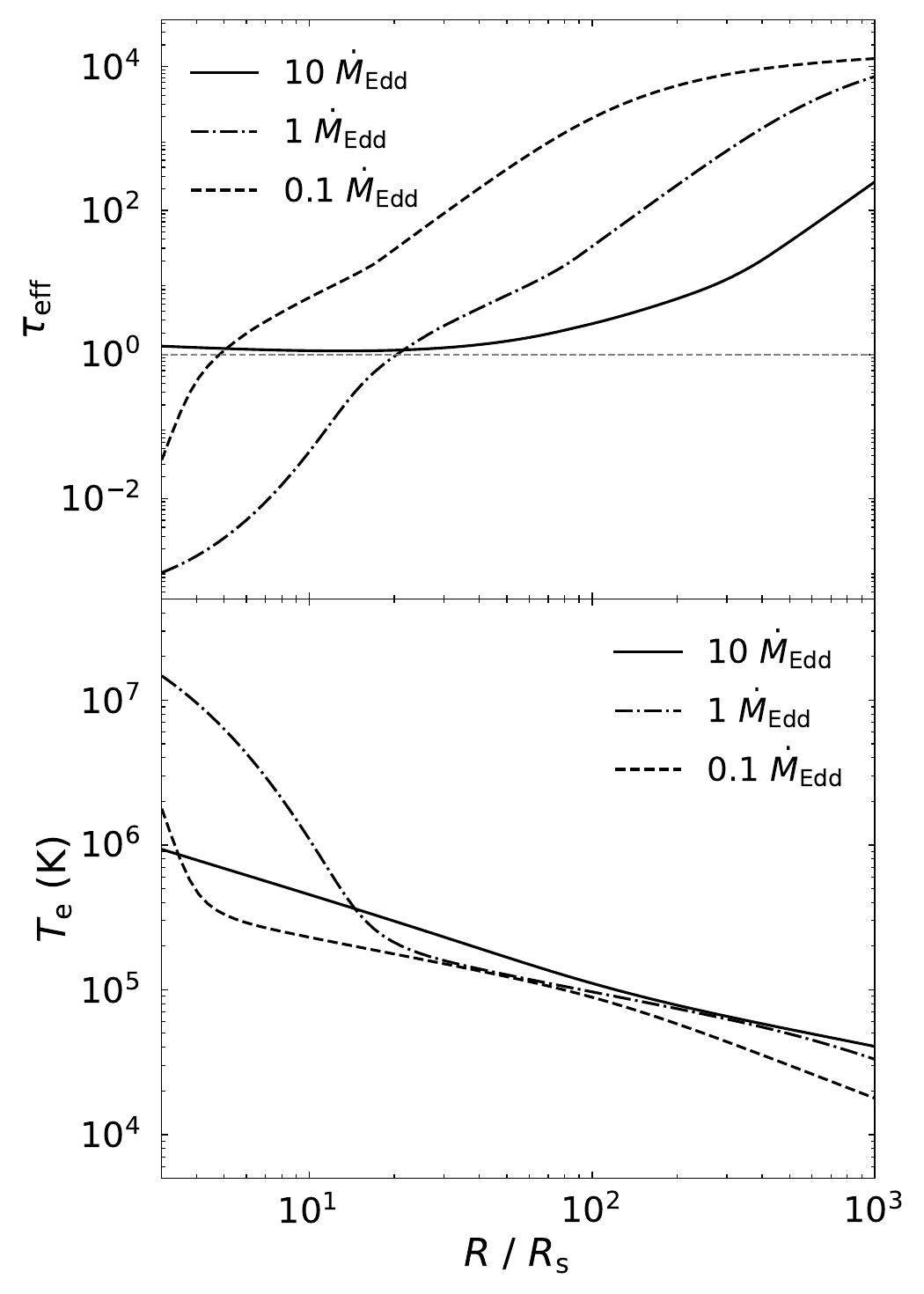}
    \caption{The radial profiles of effective optical depth (upper panel) and mid-plane temperature (lower panel) for $m = 10^8$, $\alpha=0.1$ and $p_\mathrm{m}=p_\mathrm{g}$, at $\dot{m}=0.1$ (dashed line), 1 (dash-dotted line), and 10 (solid line).}
    \label{fig:radial}
\end{figure}

Figure \ref{fig:radial} plots the radial profiles of effective optical depth and mid-plane temperature of the accretion flows for accretion rates of 0.1, 1, and 10 $\dot{M}_\mathrm{Edd}$. The grey dashed line for $\tau_{\rm eff}=1$ divides the radial region into effectively optically thin and thick regions. As it can be seen from the upper panel, the optically thin region is very narrow for $\dot m =0.1$ (dashed line), and extends to $\sim 20\ R_{\rm S}$ for $\dot m =1$ (dot-dashed line), and disappears for $\dot m =10$ (solid line) as a consequence of photon-trapping. The mid-plane temperature in the innermost region is $T\sim 10^7$ K for $\dot m=1$ and drops back to a slim disc temperature. Comparing the temperature and extension of the effectively optically thin for $\dot m=0.1$ and $\dot m=1$, one can see that the temperature at the innermost stable circular orbit (ISCO) is higher for a larger effectively optically thin region.

\subsection{Parameter space for the effectively optically thin flows} 
\label{subsec:zone}

\begin{figure}
\centering 
    \includegraphics[width=\columnwidth]{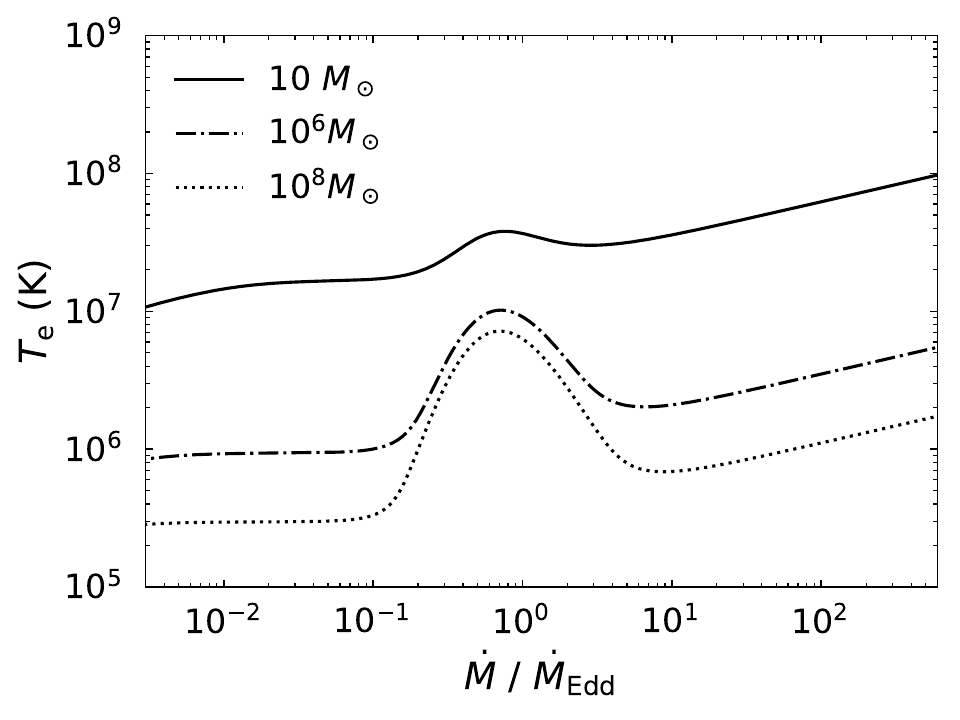}
    \caption{The mid-plane temperature as a function of mass accretion rate at the distance of 5$R_\mathrm{S}$ for different black hole masses, where $\alpha=0.1$, and $p_\mathrm{m}=p_\mathrm{g}$ are adopted. The results for $m = 10$, $10^6$ and $10^8$ are represented by solid, dot-dashed, and dotted lines, respectively.}
    \label{fig:m}
\end{figure}

The effectively optically thin accretion flow occurs at the radiation pressure dominant branch where the relation between the surface density and accretion rate of SSD is inverted, leading to an increase of the temperature so as to produce sufficient emission in balance with the accretion energy. As it is well known that the radiation pressure becomes dominant at higher $\dot m$ for smaller black holes, it is expected the transition from SSD to the effectively optically thin flow at a higher $\dot m$ for stellar mass black holes, consequently, the effectively optically thin flows exist in a narrow range of accretion rates. Figure \ref{fig:m} displays such a trend, indicated by the mid-plane temperature for 10, $10^6$, and $10^8\ M_\odot$ black holes.

\begin{figure}
\centering 
    \includegraphics[width=\columnwidth]{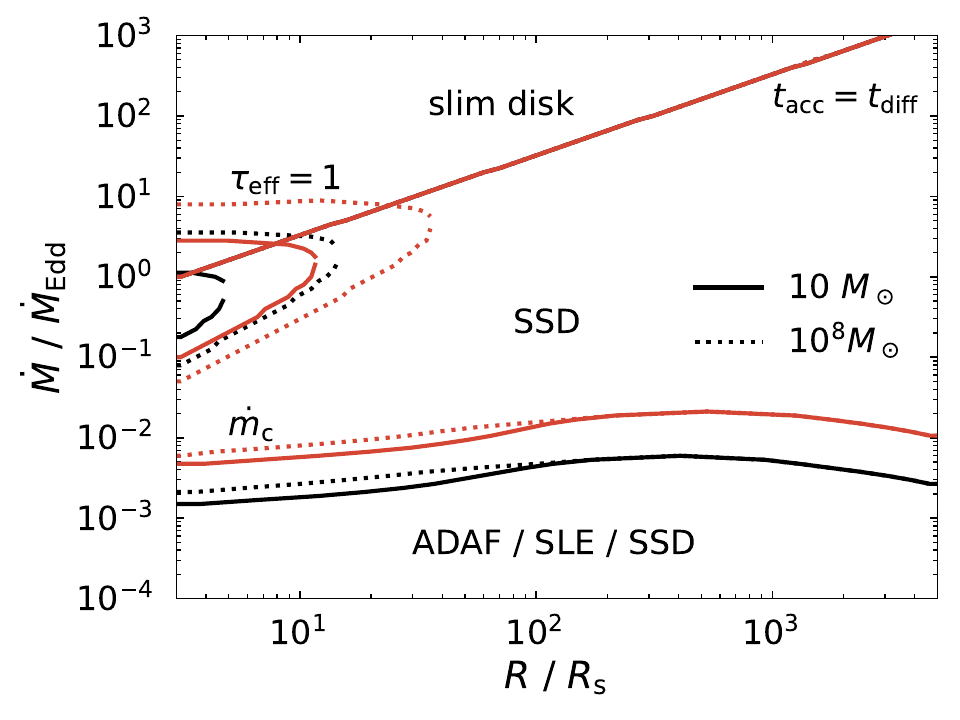}
    \caption{The parameter spaces in the $\dot{m}-r$ plane of the ADAF, the SLE, the SSD, the slim disc, and the effectively optically thin accretion flow for $m=10$ (solid lines) and 10$^8$ (dotted lines), $\alpha=0.05$ (black) and 0.1 (red), and $p_\mathrm{m}=p_\mathrm{g}$.}
    \label{fig:zone}
\end{figure}

To delineate the parameter space, we plot in Figure \ref{fig:zone} the region for the effectively optically thin accretion flow, the ADAF, the SLE, the SSD, and the slim disc in the  $\dot{M}-R$ plane. The line type reveals the effect of black hole mass, solid lines are for $ m=10$ and dotted lines for $m=10^8$; The line color reveals the effect of viscosity parameter, red for the standard viscous parameter $\alpha=0.1$ and black for $\alpha=0.05$. The lower, red curve represents the critical accretion rate of ADAF, $\dot{m}_\mathrm{c}$, indicating that the ADAF can only exist below it. The upper, straight line marked by $t_{\rm acc}=t_{\rm diff}$ corresponds to the case of accretion time equal to photon diffusion time, above which photon-trapping occurs and the accretion is via a slim disc. Between these two curves, there is only SSD. Thus, the two (red solid) lines divide the region for the existence of the four classic solutions in the $\dot{M}-R$ plane. However, we find that the region inside the red, solid parabola-like curve shown on the left of Figure \ref{fig:zone} is occupied by the effectively optically thin solution rather than the SSD or slim disc. This region of the accretion flow is at small distances with accretion rates around the boundary line between the SSD and slim disc. As the inner region of the accretion flow is the most efficient radiation region, it plays an important role in affecting the radiation spectrum considering its distinct temperature and optical depth, even though such a flow exists only in a small region in $\dot{M}-R$ space. 

Because of this, we study the occurrence of the effectively optically thin flow in dependence on the black hole mass and viscosity parameter. It is found that in a supper-massive black hole, it occurs at a larger range of accretion rate and distance than that in a stellar mass black hole, indicated by the red dotted curve and the red solid curve in Figure \ref{fig:zone}. This further confirms the results from supermassive black holes shown in Figure \ref{fig:m}. More importantly, it is found that viscosity plays a key role in the occurrence of the effectively optically thin accretion flow. As shown in Figure \ref{fig:zone}, for $\alpha=0.05$ (black), the parameter space for its existence is very narrow compared with that for $\alpha=0.1$ (red). The larger viscosity parameter means more efficient angular momentum transfer, which leads to higher radial velocity ($\upsilon\propto\alpha$) and lower surface density ($\dot M \propto \Sigma \upsilon$). Thus, the radiation pressure starts to dominate over gas pressure at a lower accretion rate or larger distances for a larger $\alpha$, leading to a larger parameter space for the existence of effectively optically thin flow. 

\begin{figure*}
\centering 
    \includegraphics[width=\textwidth]{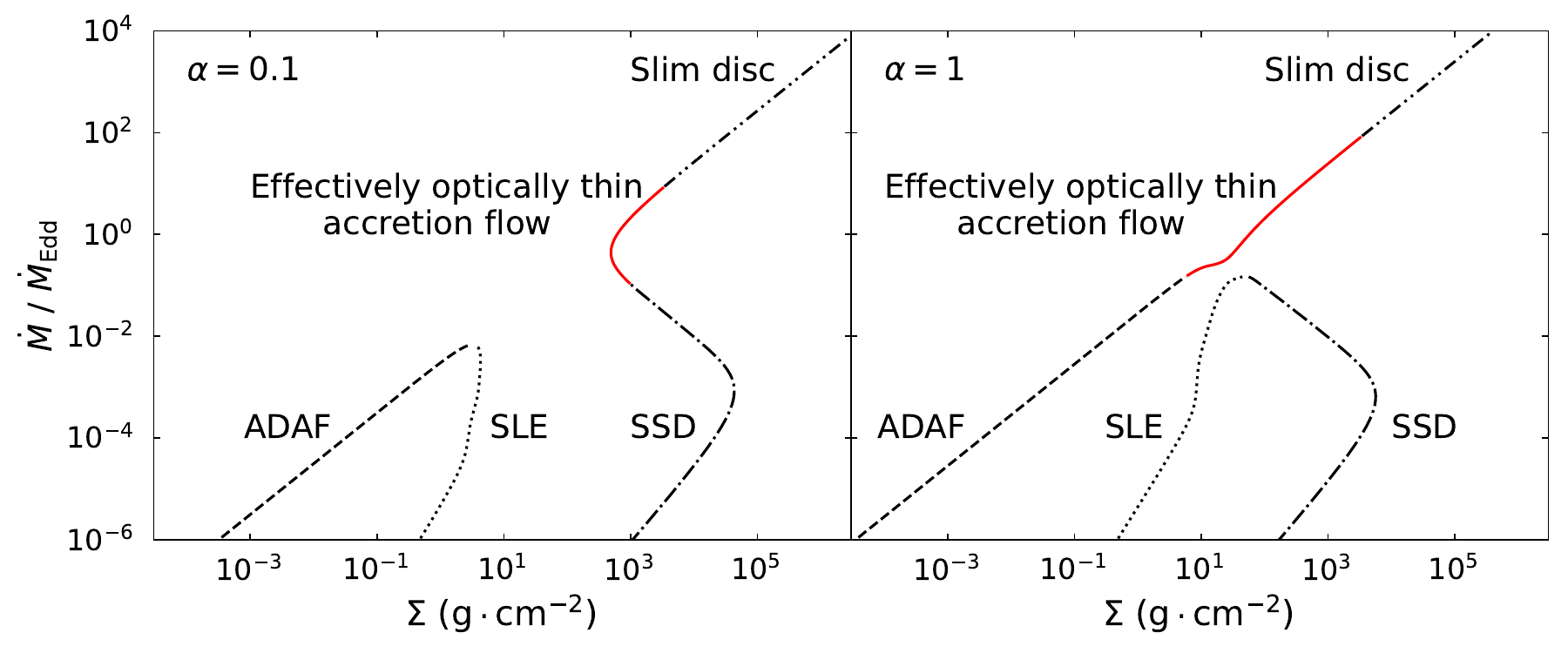}
    \caption{The generalized solutions in the $\dot m-\Sigma$ plane at the distance of 5 $R_\mathrm{S}$ for $m = 10^8$, $p_\mathrm{m}=p_\mathrm{g}$, $\alpha=0.1$ (the left panel) and $\alpha=1$ (the right panel). The effectively optically thin accretion flow (solid lines), ADAF (dashed lines), SLE (dotted lines), SSD (dash-dotted lines), and slim disc (dash-dot-dotted lines) are present.}
    \label{fig:s_curves}
\end{figure*}

Accretion flow solutions have been shown schematically on a local $\dot m-\Sigma$ plane for different viscosity parameters (Fig. 11.1 and Fig. 11.2 of \citealp{Frank_2002}, see also \citealp{Chen_1995,Bjoernsson_1996}). For comparison we display our results from detailed computations for a typical viscosity parameter $\alpha=0.1$ and an extremely large viscosity $\alpha=1$. As Figure \ref{fig:s_curves} shows, our unified solutions confirm the anticipation of \citet{Frank_2002}. Specifically,  for normal value of the viscosity parameter there is smooth connection (presented by the red, effectively optically thin curve in the left panel of \ref{fig:s_curves}) between the radiation-pressure-dominated thin disc and the slim disc, the S-curve is well separate from the ADAF-SLE solutions; With increase of $\alpha$ the ADAF-SLE branch and the S-curve move toward each other and eventually intersect for a particular value of $\alpha$, thereby the optically thin ADAF directly connects to the effectively optically thin flows with increase of accretion rate (see the right panel of Figure \ref{fig:s_curves}). The critical viscosity parameter for insection can be obtained from our computations, which is found $\alpha_\mathrm{cirt}\approx0.9$ for $m=10^8$ and $p_\mathrm{m}=p_\mathrm{g}$ at the distance of 5 $R_\mathrm{S}$. This study reveals the physical bridge for an optically thin ADAF to transfer to an optically thick slim disc for extremely large viscosity, that is,  the effectively optically thin flow.

To summarize, the effectively optically thin solution, if exists, should be in the innermost region with accretion rates around the Eddington value. Whether the effectively optically thin solution exists and to how large extent it distinguishes from the SSD or slim disc depends strongly on the viscosity parameter and the black hole mass. A large viscosity parameter in an accretion flow around a supermassive black hole makes it essential for the occurrence of the effectively optically thin solution. The effect of viscosity and black hole mass on the effectively optically thin accretion flows is demonstrated by the detailed spectra for different parameters in the following subsection.

\subsection{Emergent spectra} \label{subsec:spec}

\begin{figure*}
	\includegraphics[width=\textwidth]{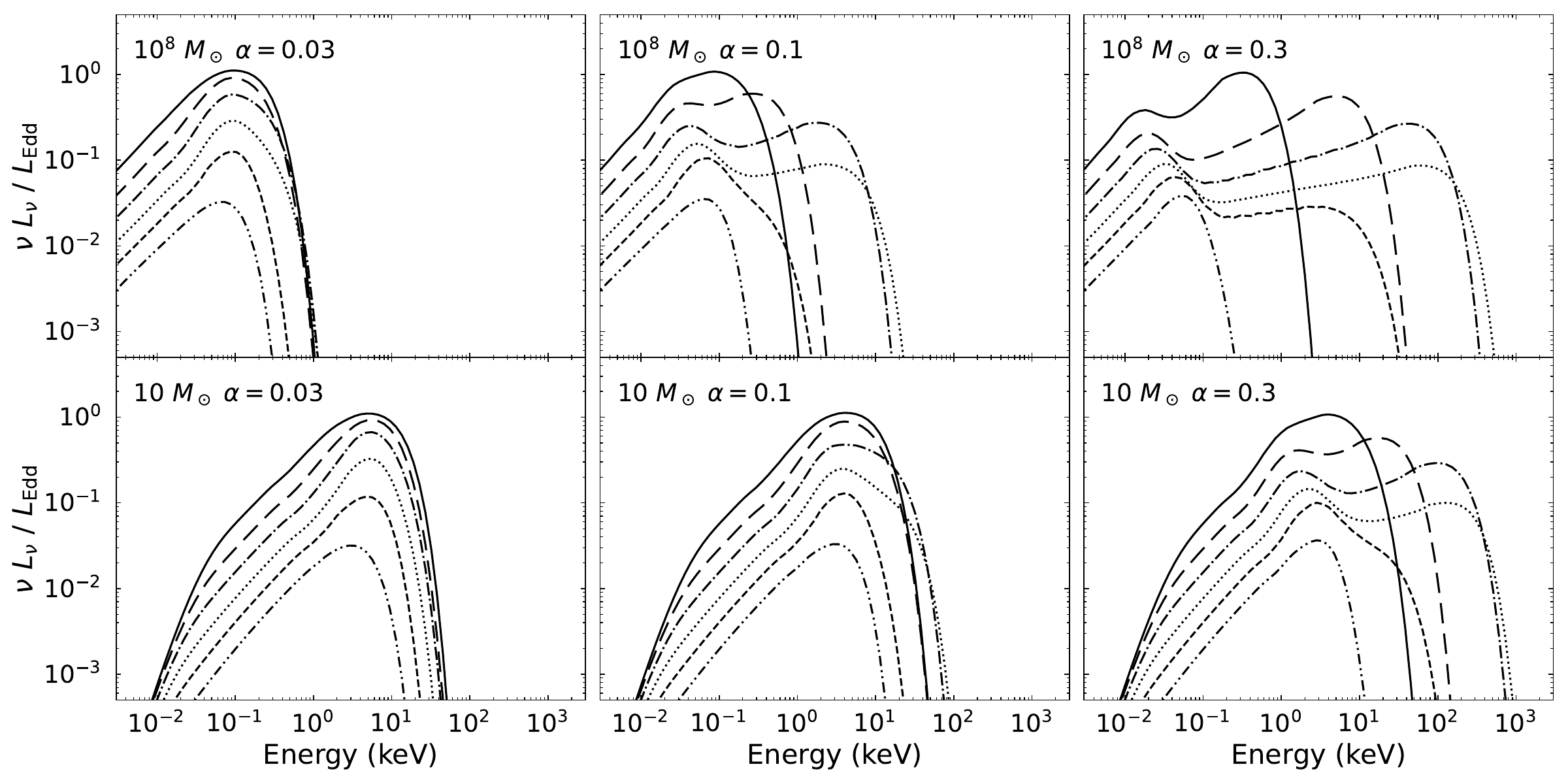}
    \caption{The spectra for $m=10^8$ (upper panels) and 10 (lower panels), $\alpha=0.03$ (left panels), 0.1 (middle panels), and 0.3 (right panels) under the assumption of equipartion pressure, $p_\mathrm{m}=p_\mathrm{g}$. In each panel the  spectra for  accretion rate $\dot{m}=0.03$, 0.1, 0.3, 1, 3 and 10 are marked in dash-dot-dotted, short dashed, dotted, solid, dash-dotted, and long dashed lines, respectively.}
    \label{fig:spec}
\end{figure*}

Figure \ref{fig:spec} plots the spectra for accretion rates ranging from sub- to super-Eddington rates. The upper panels are for a $10^8 M_\odot$ black hole and the lower panels are for a $10 M_\odot$ black hole. The viscosity parameter is set to 0.03, 0.1, and 0.3 in the left, middle, and right panels, respectively.

For the 10$^8$ $M_\odot$ black hole and standard viscosity parameter $\alpha=0.1$, the spectra are shown in the upper, middle panel of Figure \ref{fig:spec}. At $\dot{m}=0.03$ the spectrum is a typical multi-color blackbody. With the increase in accretion rate, an effectively optically thin region appears and expands outwards, and the temperature in the innermost region rises accordingly. When the accretion rate is sufficiently high, Comptonization plays a key role in radiative cooling, stopping the rising of temperature, and the effectively optically thin region shrinks and eventually recedes to ISCO, thereby the accretion flow is replaced by a slim disc (see Figure \ref{fig:radial}). This leads to the spectral variation with the accretion rate as represented by the upper, middle panel of Figure \ref{fig:spec}. The radiation from the effectively optically thin region is a multi-color Wien spectrum for the quasi-saturate Compton scattering. With the increase in accretion rate, this Wien component extends to higher frequency and higher luminosity, forming a Wien bump at around a few keV when more photons are saturated at the energy of $3kT_{e}$. When it comes to the super-Eddington stage, the cut-off energy moves back to the low frequency with the shrinking of the effectively optically thin region. Finally, the Wien component merges with the multi-color blackbody as the inner region transits to the slim disc. While the hard radiations from the inner effectively optically thin region extend to high frequency, the multi-color blackbody component from the outer SSD moves slightly to low frequency as a consequence of combining the shift of disc inner radius and the accretion rate. The total radiations form a spectrum with a typical multi-color blackbody at a few 10 eV and a power-law component in X-rays, with a slope and cut-off frequency depending on the accretion rate.   

Since the occurrence of the effectively optically thin accretion is sensitive to the viscosity parameter, we present the spectra for $\alpha=0.03$ and $\alpha=0.3$ for comparison. As shown in the upper, left panel of Figure \ref{fig:spec}, the power-law component almost disappears for $\alpha=0.03$. Instead, a narrow effectively optically thin region produces an additional component, which is more like a blackbody, constantly locating at around 0.1keV for different accretion rates. On the other hand, for $\alpha=0.3$  the power spectra (see the upper, right panel of Figure \ref{fig:spec}) can extend to a few hundred keV. The high energy cutoff strongly depends on the accretion rate. The large effectively optically thin region dominates the bolometric luminosity for accretion rate in the range of $\dot{m}=0.3-1$.

For stellar mass black holes, the effectively optically thin accretion flow occupies a narrow parameter space. In most cases, it does not exist. For the standard viscosity, $\alpha=0.1$, the effectively optically thin accretion flow appears for near- or super-Eddington rates, and its effect on the spectra is not much different from color correction. Only for large viscosity (say $\alpha=0.3$) could it become strong. The results are displayed in the lower panels of Figure \ref{fig:spec}. 
 
The occurrence of effectively optically thin accretion flow near ISCO in supermassive black holes is particularly interesting. The occurrence of a narrow, moderate temperature region of the effectively optically thin region for a small $\alpha=0.03$ provides an extra component to the thin disc emission, which may be relevant to soft X-ray excess; The strong, hot, effectively optically thin accretion flow for $\alpha=0.1-0.3$ can produce luminous hard X-ray emissions, much stronger than the traditional hot flows, such as the ADAF or corona. This might provide a promising mechanism for the strong X-ray emissions as observed in AGN, which is far too luminous for a corona to produce unless extra unknown heating is assumed for the corona according to the classic picture of accretion flows. We discuss these effects as well as the constraints on the viscous parameter in Section \ref{sec: dis-obs}.

\subsection{Stability} \label{subsec:stable}

Since the effectively optically thin accretion flow is supported by radiation pressure, it has been speculated to be unstable \citep[e.g.,][]{Callahan_1977,Sakimoto1981} and did not attract much attention. However, we note that in a small effective optical depth gas the radiation pressure, though it still dominates over the gas pressure, is much less than $aT^4$ (see Eq.\ref{eq:EoS}), as also pointed out by \citet{Artemova_2006}. This could alleviate the unstable feature and establish a stable bridge between the SSD and slim disc with the increase of accretion rate. In the following, we study the stability of the effectively optically thin accretion flow in detail.

In our previous work \citepalias{Liu2025}, we have developed a method of thermal stability analysis, which self-consistently includes the effect of advection. For radiation pressure-supported accretion flow, the vertically integrated heating and radiative cooling rates $Q^+$ and $Q^-$, the scale height, and density are estimated as follows,
\begin{equation}
    \begin{aligned}
         &Q^+\approx1.7\times10^{27}c_2^2\Xi m^{-1}\dot{m}r^{-3}\ \text{erg}\ \text{cm}^{-2}\ \text{s}^{-1},\\
         &Q^-\approx4.0\times10^{26}\left(1-\varepsilon_\mathrm{sca}\right)\left(c^\prime f\Xi\right)^{1/2}m^{-1}r^{-2}\ \text{erg}\ \text{cm}^{-2}\ \text{s}^{-1},\\
         &H\approx5.7\times10^5\left(c^\prime f\Xi\right)^{1/2}mr\ \text{cm},\\
        &\rho\approx1.4\times10^{-5}\left(c^\prime f\Xi\right)^{-3/2}\left(\alpha m\right)^{-1}\dot{m}r^{-3/2}\ \text{g}\ \text{cm}^{-3},
    \end{aligned}
\end{equation}
where $c_2\equiv\Omega/\Omega_\mathrm{K}$, $\Xi\equiv\left(q_\mathrm{vis}+q_\mathrm{c}\right)/q_\mathrm{vis}$, $c^\prime\equiv c_2^2\left(\Gamma_3-1\right)$, and $1-\varepsilon_\mathrm{sca}\equiv\tau_\mathrm{es}/\tau$. For the radiation-pressure-dominated case, $1-\varepsilon_\mathrm{sca}\approx1$, $\Xi\approx2/(2-f)$ and $c^\prime\approx c_2^2/3$ at the order of unity. 

We write the cooling rate as $Q^-=2F_\mathrm{rad}\approx2\eta_\mathrm{br}q_\mathrm{br}H$ and employ $q_\mathrm{br}=5.43\times10^{20}\rho^2T^{1/2}\ \text{erg}\ \text{cm}^{-3}\ \text{s}^{-1}$ \citep{Rybicki_Lightman_1979} to express the mid-plane temperature as follows,
\begin{equation}\label{eq:T}
    T\approx1.2\times10^{19}\eta_\mathrm{br}^{-2}\left(c^\prime f\Xi\right)^6\alpha^4 \dot{m}^{-4}\ \text{K}.
\end{equation}
Since the Compton scattering is almost saturated, the average luminosity enhancement factor $\eta\approx3\theta_\mathrm{e}/x$ \citep{Dermer_1991,Narayan_Yi_1995}. Therefore, the integrated factor from the self-absorption frequency $x_\mathrm{c}$ to electron temperature $\theta_\mathrm{e}$ for bremsstrahlung  $\eta_\mathrm{br}\approx3\ln{\left(\theta_\mathrm{e}/x_\mathrm{c}\right)}$ is non-sensitive to the temperature according to Eq.(\ref{eq:comp}). 

\begin{figure}
    \centering 
    \includegraphics[width=\columnwidth]{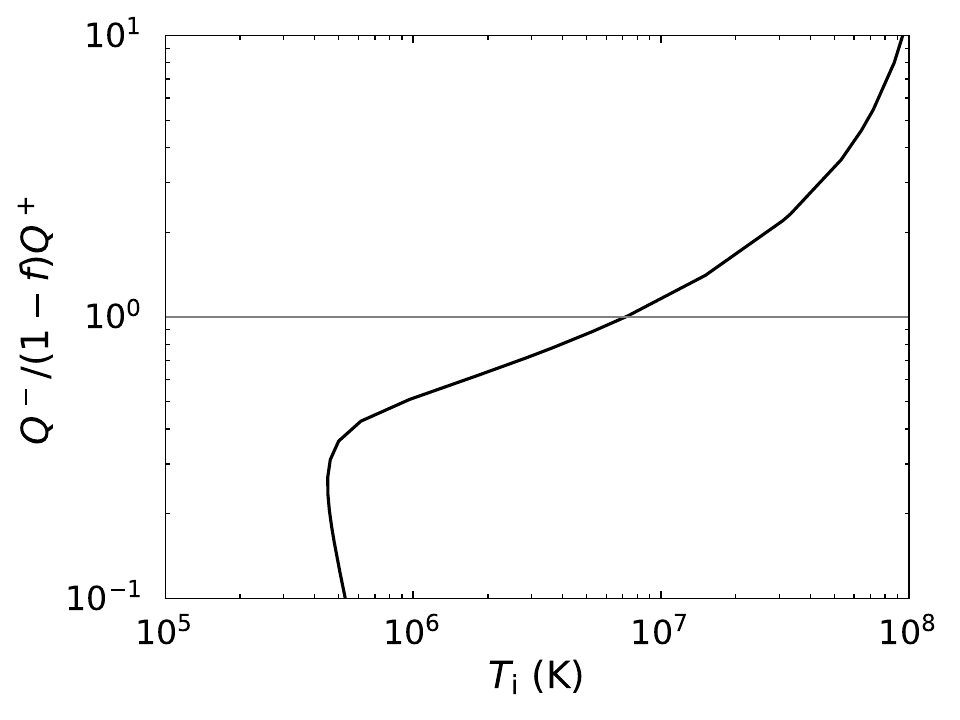}
    \caption{The ratio between the cooling rate and heating rate as a function of ion temperature for $m=10^8$, $\dot{m}=0.8$, $\alpha=0.1$, $p_\mathrm{m}=p_\mathrm{g}$, and $R=5 R_\mathrm{S}$. The positive slope of ratio around the equilibrium solution at $Q^-/(1-f)Q^+=1$ indicates the solution is thermal stable.}
    \label{fig:stable}
\end{figure}

With Eq.(\ref{eq:T}), the effectively optically thin accretion flow satisfies the stability criterion (\ref{eq:stable}),
\begin{equation}
   \frac{\text d\left(1-f\right)Q^+}{\text dT}<\frac{\text dQ^-}{\text dT}.
   \label{eq:stable}
\end{equation}
The thermal stability is also numerically verified by the variation of $Q^-/(1-f)Q^+$ with temperature, as shown in Figure \ref{fig:stable} (the method can be seen in \citetalias{Liu2025}). The point for $Q^-/(1-f)Q^+=1$ corresponds to the thermal equilibrium solution, around which the slope of the $Q^-/(1-f)Q^+$ in the logarithmic graph is positive, indicating $\text d\ln Q^-/\text d\ln T>\text d\ln\left(1-f\right)Q^+/\text d\ln T$. Thus,  any thermal perturbation on the accretion flow is suppressed. 

We perform the viscosity stability analysis following \citet{Kato_2008}. The criterion of viscous instability is $\partial T_\mathrm{r,\phi}/\partial \Sigma>0$, where $T_\mathrm{r,\phi}$ is the vertically integrated torque. Considering the perturbation on the accretion flow around the equilibrium state, we can decompose the torque, pressure, mid-plane temperature, surface density, and height as the equilibrium part and perturbation part, i.e., $T_\mathrm{r,\phi} = T_\mathrm{r,\phi,0} + T_\mathrm{r,\phi,1}$, $p=p_0+p_1$, $T=T_0+T_1$, $\Sigma = \Sigma_0 + \Sigma_1$ and $H=H_0+H_1$, where the subscripts 0 and 1 denote the equilibrium and perturbed quantities, respectively. We then examine the value of  $\partial T_\mathrm{r,\phi}/\partial \Sigma \approx T_\mathrm{r,\phi,1}/\Sigma_1$ is positive or negative. The perturbations of $\Omega$ and $\eta_\mathrm{br}$ are weak and can be ignored. 

We need three relations to eliminate perturbations of pressure, temperature, and height. The vertical hydrostatic equilibrium equation gives the following relation,
\begin{equation}
    \frac{p_1}{p_0} = \frac{\Sigma_1}{\Sigma_0} + \frac{H_1}{H_0}.
    \label{eq:vis_1}
\end{equation}

The second relation is derived from the expression of radiation pressure, i.e., $p_\mathrm{r}\propto F_\mathrm{rad}\Sigma\propto\eta_\mathrm{br}\Sigma^3H^{-1}T^{1/2}$. For the radiation-pressure-supported accretion flow, it can be written as follows,
\begin{equation}
    \frac{p_1}{p_0}\approx3\frac{\Sigma_1}{\Sigma_0}-\frac{H_1}{H_0}+\frac{1}{2}\frac{T_1}{T}.
    \label{eq:vis_2}
\end{equation}

The balance of energy (\ref{eq:energy_total}) gives the third relation. The viscous heating satisfies $q_\mathrm{vis}\propto\alpha p\Omega$. The advection and compression work are expressed as $q_\mathrm{int}\propto q_\mathrm{c}\propto p\upsilon\propto p\Sigma^{-1}$ with the continuity equation (\ref{eq:continuity}). The radiative cooling rate $q_\mathrm{rad}\propto F_\mathrm{rad}H^{-1}\propto\eta_\mathrm{br}\Sigma^2H^{-2}T^{1/2}$ via the expression of radiation pressure as mentioned above. We eliminate the $q_\mathrm{vis}$, $q_\mathrm{int}$, $q_\mathrm{c}$, and $q_\mathrm{rad}$ through $f$ and $\Xi$, i.e, $q_\mathrm{int}=f\Xi q_\mathrm{vis}$, $q_\mathrm{rad}=(1-f)\Xi q_\mathrm{vis}$, and $q_\mathrm{c}=(\Xi-1)q_\mathrm{vis}$. These give the third relation,
\begin{equation}
    \frac{p_1}{p_0}=\frac{4-5f}{2\left(1-f\right)}\frac{\Sigma_1}{\Sigma_0}-2\frac{H_1}{H_0}+\frac{1}{2}\frac{T_1}{T}.
    \label{eq:vis_3}
\end{equation}

Finally, we use $\Sigma_1/\Sigma_0$ to express $p_1/p_0$, $H_1/H_0$ and $T_1/T_0$ in Eq.(\ref{eq:vis_1}), (\ref{eq:vis_2}) and (\ref{eq:vis_3}). From the definition of torque ($T_\mathrm{r,\phi}\propto\alpha pH$), we can obtain the relation between torque and surface density,
\begin{equation}
    \frac{T_{r\phi,1}}{T_{r\phi,0}}=\frac{p_1}{p_0}+\frac{H_1}{H_0}=-\frac{1}{1-f}\frac{\Sigma_1}{\Sigma_0}.
\end{equation}
Since the torque $T_{r\phi,0}<0$, the ratio $T_{r\phi,1}/\Sigma_1$ is positive, corresponding to $\partial T_\mathrm{r,\phi}/\partial \Sigma>0$. This effectively optically thin accretion flow is viscously unstable.

Therefore, we conclude that the effectively optically thin accretion flow is thermally stable but viscously unstable.

\section{Discussion} \label{sec:discussion}

\subsection{The absorption edge of the soft photons for Comptonization}

\begin{figure}
    \centering 
    \includegraphics[width=\columnwidth]{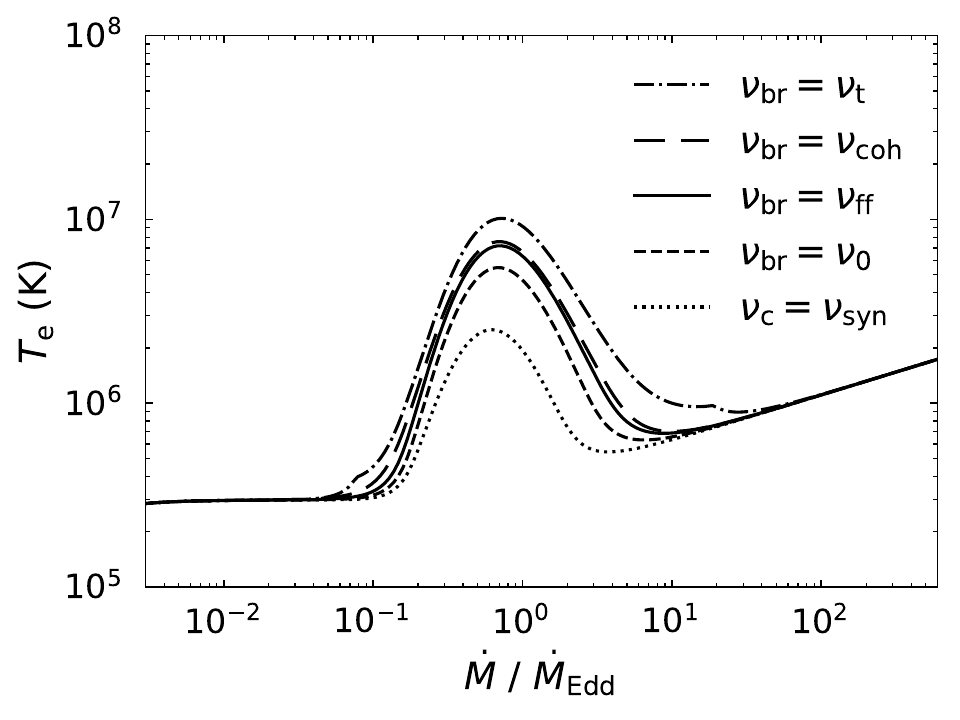}
    \caption{The effect of absorption edge on the mid-plane temperatures $T_{\rm e}(\dot M/\dot M_{\rm Edd})$ for $m=10^8$, $\alpha=0.1$, $p_\mathrm{m}=p_\mathrm{g}$, and $R=5 R_\mathrm{S}$.}
    \label{fig:nu_cut}
\end{figure}

In the effectively optically thin accretion flow, Comptonization becomes important in radiative cooling. The cooling rate strongly depends on the number of soft photons in the quasi-saturated scattering. Thus, the absorption edge of the bremsstrahlung and synchrotron as soft photons becomes crucial in evaluating the Compton cooling. A higher absorption edge means fewer soft photons, leading to a weaker Comptonization and hence a higher temperature as a consequence of energy balance. To understand this influence, we perform the calculations for choosing different characteristic frequencies as the absorption edge, that is, the critical frequency for synchrotron $(\nu_\mathrm{syn}$) and for bremsstrahlung ($\nu_\mathrm{ff}$) to become optically thick, the effectively optically thick frequency ($\nu_\mathrm{t}$), the frequency for equal scattering and absorption coefficients ($\nu_0$), and the frequency for the Compton $y$-parameter equal to unity ($\nu_\mathrm{coh}$). At accretion rates sufficiently high so that the scatterings is near saturation, we can derive the relation $\nu_\mathrm{syn}<\nu_0<\nu_\mathrm{ff}\approx\nu_\mathrm{coh}<\nu_\mathrm{t}$. Since the synchrotron radiation rate turns out to be weaker than the bremsstrahlung rate, the appropriate absorption edge should be either $\nu_\mathrm{ff}$ or $\nu_\mathrm{t}$. Adopting $\nu_\mathrm{ff}$ can over-estimate the seed photons, while adopting $\nu_\mathrm{t}$ can slightly under-estimate the seed photons to be practically scattered. $\nu_\mathrm{0}$ is not better than $\nu_\mathrm{ff}$ since it is not directly related to the  absorption. The Comptonization is well included, see Eq.(\ref{eq:eta}), no matter whether Compton $y$-parameter is larger or less than 1, so it is not necessary to choose $\nu_\mathrm{coh}$, nevertheless, it can give a moderate description for the seed photons as it is in between $\nu_\mathrm{ff}$ and $\nu_\mathrm{t}$. In Figure \ref{fig:nu_cut} we show the effect of all possible absorption edges on the temperature, where the true temperature should be between the curves for $\nu_\mathrm{ff}$ and $\nu_\mathrm{t}$. With the increase of the chosen absorption edge, the Comptonization rate decreases, thus, the temperature has to increase in order to radiate away the released gravitational energy. Figure \ref{fig:nu_cut} shows the temperature difference obtained for absorption edge from $\nu_\mathrm{ff}$ to $\nu_\mathrm{t}$ is not large. Therefore, our results obtained by adopting the absorption edge $\nu_\mathrm{br}=\nu_\mathrm{ff}$ are a good approximation.

\subsection{Observational Implication}\label{sec: dis-obs}

With the effectively optically thin accretion flow, we reveal a new spectral feature, i.e., a multi-color blackbody at low frequencies combined with a multi-color Wien spectrum at high frequencies, significantly different from the multi-color blackbody for SSD and slim disc. This feature exists at the accretion rates around $\dot{M}_\mathrm{Edd}$ and large viscosity parameters, e.g., $\alpha>0.03$ for $m=10^8$ and $\alpha>0.1$ for $m=10$. The Wien component lies on the X-ray band and is sensitive to the viscosity parameter. Therefore, our results help to constrain the viscosity parameter by comparing the observational spectra with the theoretical prediction. In addition, it can help to understand some unusual spectra and phenomena in accreting systems. 

\begin{figure*}
\centering 
    \includegraphics[width=\textwidth]{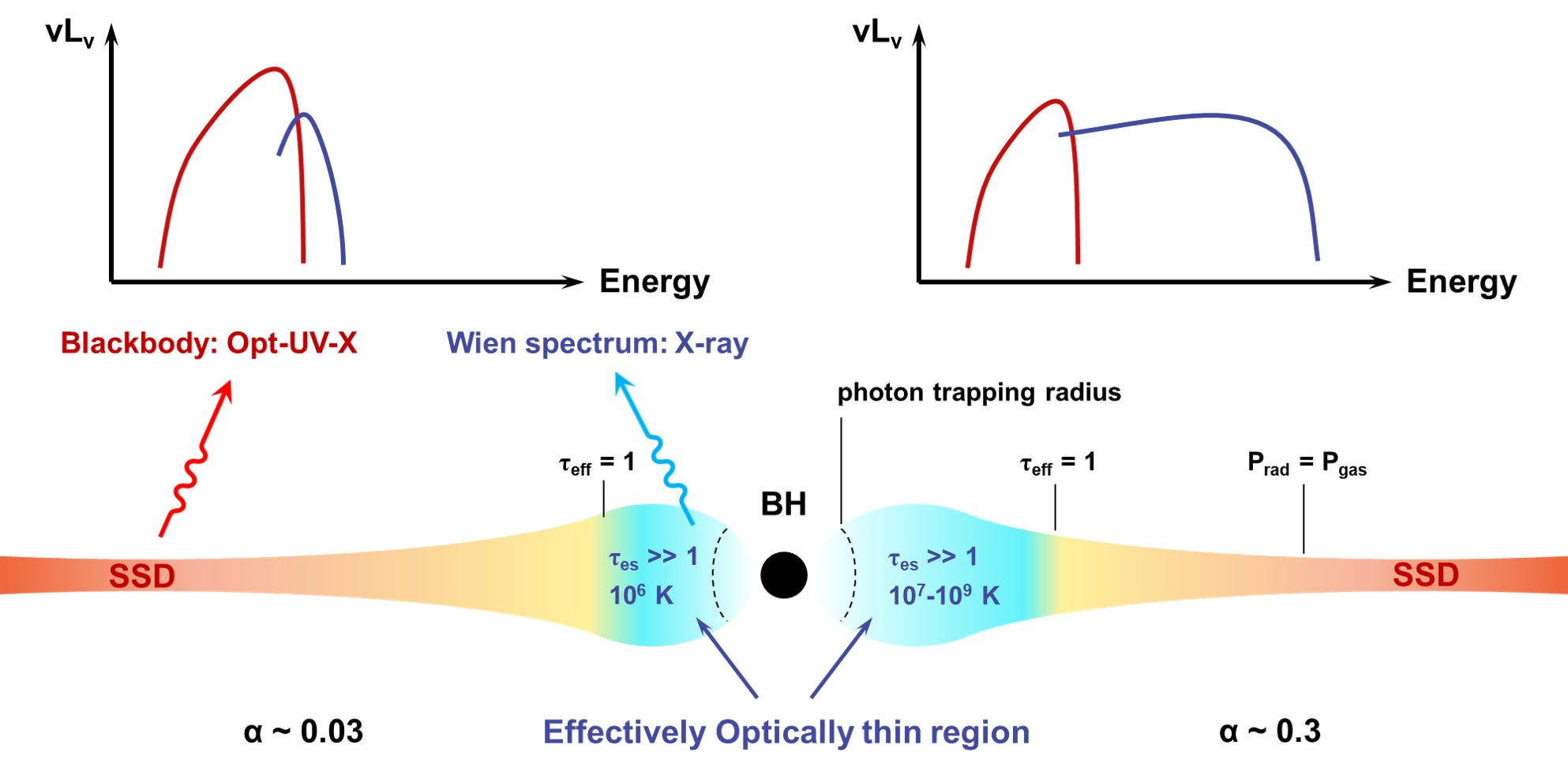}
    \caption{A schematic description  of accretion flows with the effectively optically thin region for $\alpha\sim 0.03$ (left) and $\alpha\sim 0.3$ (right). }
    \label{fig:geo}
\end{figure*}

\citet{King_2007} estimated $\alpha$ as 0.2-0.4 for XRBs and  0.01-0.03 for AGNs, via the observed variability. With a consistent viscosity, $\alpha=0.3$, the lower-right panel of Figure \ref{fig:spec} shows that the spectra for XRBs at high accretion rates would have a power-law component extending to several hundred keV with the photon index not less than two. It seems to be consistent with the very high state in the outburst of XRBs \citep[see e.g.,][]{Remillard_McClintock_2006}. For AGNs, the spectra from accretion flows with a consistent viscosity of $\alpha=0.03$ \citep{King_2007} are shown in the upper, left panel. With the occurrence of a small effectively optically thin region, the radiations peak around 0.1keV at $\dot m\sim 0.1$. For higher viscosity, e.g., $\alpha=0.1$, the spectra largely deviate from the average observational spectrum of AGN, constraining the viscosity parameter to values around $0.03$, in agreement with \citet{King_2007}. We summarize these results into a schematic diagram, as Figure \ref{fig:geo} shows.

It is interesting to note that the occurrence of effectively optically thin region provides an additional blackbody/Wien component to AGN spectra for the constrained $\alpha=0.03$. This component peaks at a frequency slightly higher than 0.1 keV, which does not depend sensitively on the accretion rate ranging from 0.1 to 10 (see the upper, left panel of Figure \ref{fig:spec}). Such a feature could provide a physical interpretation for the origin of soft X-ray excess, which has been under debates including the warm corona \citep[e.g.,][]{Magdziarz_1998,Done_2012,Petrucci_2013,Petrucci_2018} and blurred ionized reflection \citep[e.g.,][]{Gierlinski_Done_2004,Crummy_2006,Zoghbi_2008,Walton_2013}. In particular, for very luminous AGNs with accretion rates around Eddington value, such as the narrow line Seyfert 1 galaxies and the changing-look AGN 1ES 1927+654, an extra blackbody component must be added to the disc and corona components in fitting the observational spectra. This component is too hot to compare with the thin or slim disc, but too cool to compare with a typical corona. Thus it was assumed to originate from an overheated disc \citep[e.g.,][]{Li2024a,Li2024b} or a warm corona \citep[e.g.,][]{Jin2017a,Jin2017b}. With the existence of the effectively optically thin flow at the corresponding accretion rates, the blackbody component and the optical/UV to X-ray spectra are expected to be naturally explained by the physically structured accretion flows, that is, the outer thin disc, the innermost effectively optically thin flow and the corona, for which the geometry and spectrum are schematically described by the left part of Figure \ref{fig:geo}. We leave the in-depth explanations in future work.

The power-law X-ray radiation in radio-quiet AGNs is so strong that its luminosity can be comparable with the SSD component in the optical/UV band, much stronger than that in typical XRBs. What powers a corona to produce such luminous X-ray rays has puzzled theoreticians because the viscous heating in a corona is never  sufficient for the strong X-rays. If the viscosity parameter in AGN accretion flow could be as large as that in black hole X-ray binaries, the effectively optically thin accretion flow would be strong enough to interpret the power-law X-ray emissions, as can be seen in the upper, right panel of Figure \ref{fig:spec}. The typical geometry and spectrum are displayed in the right part of Figure \ref{fig:geo}.

\section{Conclusion} \label{sec:conclusion}

With the generalized solution of accretion flows developed in \citetalias{Liu2025}, we revisit a long-neglected solution - the effectively optically thin accretion flow. We find that this solution can exist in a large range of accretion rates, bridging the radiation-pressure-dominant thin disc and the slim disc. When a thin disc is radiation pressure supported, the density decreases with the increase of accretion rate while the temperature does not vary, making the innermost region effectively optically thin at a certain accretion rate. Meanwhile, the bremsstrahlung is not efficient enough to cool the viscous heating released by accretion. The gas is then heated to a high temperature so that the inverse-Compton scattering plays a role in the balance of the accretion energy. This optically thin regime exists at moderate accretion rates between the SSD and slim disc regimes and extends to a larger range of accretion rates for more massive black holes and larger viscosity parameters. Because of the large scattering optical depth, the Compton scattering is almost saturated, contributing an additional multi-color Wien component to the typical multi-color blackbody of the outer SSD region. Our stability analysis proves the effectively optically thin accretion flow thermally stable, indicating its existence in some luminous accreting systems. Compared with average AGN spectra, our generalized solution constrains the viscosity parameter to be $\alpha\sim 0.03$ and could provide an alternative interpretation on the soft X-ray excess with the existence of a narrow effectively optically thin region. The extra blackbody component shown in systems with Eddington accretion, speculated from a warm corona or overheated disc, could originate from the inner effectively optically thin region. On the other hand, if the viscous parameter could be as large as that in black hole XRBs, the effectively optically thin region might correspond to the strong X-ray emissions in AGNs. 

\section*{Acknowledgements}

We thank the anonymous referee and Erlin Qiao for their valuable comments. We acknowledge the support by the National Natural Science Foundation of China (Grants No.12333004 and 12433005). We also acknowledge the Beijing Super Cloud Center (BSCC) for providing HPC resources.

\section*{Data Availability}
The code and data underlying this article will be shared on reasonable request to the corresponding author.




\begin{thebibliography}{}
\makeatletter
\relax
\def\mn@urlcharsother{\let\do\@makeother \do\$\do\&\do\#\do\^\do\_\do\%\do\~}
\def\mn@doi{\begingroup\mn@urlcharsother \@ifnextchar [ {\mn@doi@}
  {\mn@doi@[]}}
\def\mn@doi@[#1]#2{\def\@tempa{#1}\ifx\@tempa\@empty \href
  {http://dx.doi.org/#2} {doi:#2}\else \href {http://dx.doi.org/#2} {#1}\fi
  \endgroup}
\def\mn@eprint#1#2{\mn@eprint@#1:#2::\@nil}
\def\mn@eprint@arXiv#1{\href {http://arxiv.org/abs/#1} {{\tt arXiv:#1}}}
\def\mn@eprint@dblp#1{\href {http://dblp.uni-trier.de/rec/bibtex/#1.xml}
  {dblp:#1}}
\def\mn@eprint@#1:#2:#3:#4\@nil{\def\@tempa {#1}\def\@tempb {#2}\def\@tempc
  {#3}\ifx \@tempc \@empty \let \@tempc \@tempb \let \@tempb \@tempa \fi \ifx
  \@tempb \@empty \def\@tempb {arXiv}\fi \@ifundefined
  {mn@eprint@\@tempb}{\@tempb:\@tempc}{\expandafter \expandafter \csname
  mn@eprint@\@tempb\endcsname \expandafter{\@tempc}}}

\bibitem[\protect\citeauthoryear{{Abramowicz}, {Czerny}, {Lasota}  \&
  {Szuszkiewicz}}{{Abramowicz} et~al.}{1988}]{Abramowicz_1988}
{Abramowicz} M.~A.,  {Czerny} B.,  {Lasota} J.~P.,   {Szuszkiewicz} E.,  1988,
  \mn@doi [\apj] {10.1086/166683}, \href
  {https://ui.adsabs.harvard.edu/abs/1988ApJ...332..646A} {332, 646}


\bibitem[\protect\citeauthoryear{{Artemova}, {Bisnovatyi-Kogan}, {Igumenshchev}
   \& {Novikov}}{{Artemova} et~al.}{2006}]{Artemova_2006}
{Artemova} Y.~V.,  {Bisnovatyi-Kogan} G.~S.,  {Igumenshchev} I.~V.,   {Novikov}
  I.~D.,  2006, \mn@doi [\apj] {10.1086/496964}, \href
  {https://ui.adsabs.harvard.edu/abs/2006ApJ...637..968A} {637, 968}

\bibitem[\protect\citeauthoryear{{Beloborodov}}{{Beloborodov}}{1998}]{Beloborodov_1998}
{Beloborodov} A.~M.,  1998, \mn@doi [\mnras]
  {10.1046/j.1365-8711.1998.01530.x}, \href
  {https://ui.adsabs.harvard.edu/abs/1998MNRAS.297..739B} {297, 739}

\bibitem[\protect\citeauthoryear{{Bjoernsson}, {Abramowicz}, {Chen}  \&
  {Lasota}}{{Bjoernsson} et~al.}{1996}]{Bjoernsson_1996}
{Bjoernsson} G.,  {Abramowicz} M.~A.,  {Chen} X.,   {Lasota} J.-P.,  1996,
  \mn@doi [\apj] {10.1086/177587}, \href
  {https://ui.adsabs.harvard.edu/abs/1996ApJ...467...99B} {467, 99}

\bibitem[\protect\citeauthoryear{Blundell \& Blundell}{Blundell \&
  Blundell}{2009}]{Blundell_Blundell_2009}
Blundell S.~J.,  Blundell K.~M.,  2009, {Concepts in Thermal Physics}.
Oxford University Press, \mn@doi{10.1093/acprof:oso/9780199562091.001.0001},
  \url {https://doi.org/10.1093/acprof:oso/9780199562091.001.0001}

\bibitem[\protect\citeauthoryear{{Brussaard} \& {van de Hulst}}{{Brussaard} \&
  {van de Hulst}}{1962}]{Brussaard_1962}
{Brussaard} P.~J.,  {van de Hulst} H.~C.,  1962, \mn@doi [Reviews of Modern
  Physics] {10.1103/RevModPhys.34.507}, \href
  {https://ui.adsabs.harvard.edu/abs/1962RvMP...34..507B} {34, 507}

\bibitem[\protect\citeauthoryear{{Callahan}}{{Callahan}}{1977}]{Callahan_1977}
{Callahan} P.~S.,  1977, \aap, \href
  {https://ui.adsabs.harvard.edu/abs/1977A&A....59..127C} {59, 127}

\bibitem[\protect\citeauthoryear{{Chandrasekhar}}{{Chandrasekhar}}{1967}]{Chandrasekhar_1967}
{Chandrasekhar} S.,  1967, {An introduction to the study of stellar structure}.
New York: Dover

\bibitem[\protect\citeauthoryear{{Chen} \& {Wang}}{{Chen} \&
  {Wang}}{2004}]{Chen_Wang_2004}
{Chen} L.-H.,  {Wang} J.-M.,  2004, \mn@doi [\apj] {10.1086/423416}, \href
  {https://ui.adsabs.harvard.edu/abs/2004ApJ...614..101C} {614, 101}

\bibitem[\protect\citeauthoryear{{Chen}, {Abramowicz}, {Lasota}, {Narayan}  \&
  {Yi}}{{Chen} et~al.}{1995}]{Chen_1995}
{Chen} X.,  {Abramowicz} M.~A.,  {Lasota} J.-P.,  {Narayan} R.,   {Yi} I.,
  1995, \mn@doi [\apjl] {10.1086/187836}, \href
  {https://ui.adsabs.harvard.edu/abs/1995ApJ...443L..61C} {443, L61}

\bibitem[\protect\citeauthoryear{{Clayton}}{{Clayton}}{1983}]{Clayton_1983}
{Clayton} D.~D.,  1983, {Principles of stellar evolution and nucleosynthesis}.
Chicago: University of Chicago Press

\bibitem[\protect\citeauthoryear{{Crummy}, {Fabian}, {Gallo}  \&
  {Ross}}{{Crummy} et~al.}{2006}]{Crummy_2006}
{Crummy} J.,  {Fabian} A.~C.,  {Gallo} L.,   {Ross} R.~R.,  2006, \mn@doi
  [\mnras] {10.1111/j.1365-2966.2005.09844.x}, \href
  {https://ui.adsabs.harvard.edu/abs/2006MNRAS.365.1067C} {365, 1067}

\bibitem[\protect\citeauthoryear{{Czerny} \& {Elvis}}{{Czerny} \&
  {Elvis}}{1987}]{Czerny_Elvis_1987}
{Czerny} B.,  {Elvis} M.,  1987, \mn@doi [\apj] {10.1086/165630}, \href
  {https://ui.adsabs.harvard.edu/abs/1987ApJ...321..305C} {321, 305}

\bibitem[\protect\citeauthoryear{{Dermer}, {Liang}  \& {Canfield}}{{Dermer}
  et~al.}{1991}]{Dermer_1991}
{Dermer} C.~D.,  {Liang} E.~P.,   {Canfield} E.,  1991, \mn@doi [\apj]
  {10.1086/169770}, \href
  {https://ui.adsabs.harvard.edu/abs/1991ApJ...369..410D} {369, 410}

\bibitem[\protect\citeauthoryear{{Done}, {Davis}, {Jin}, {Blaes}  \&
  {Ward}}{{Done} et~al.}{2012}]{Done_2012}
{Done} C.,  {Davis} S.~W.,  {Jin} C.,  {Blaes} O.,   {Ward} M.,  2012, \mn@doi
  [\mnras] {10.1111/j.1365-2966.2011.19779.x}, \href
  {https://ui.adsabs.harvard.edu/abs/2012MNRAS.420.1848D} {420, 1848}

\bibitem[\protect\citeauthoryear{{Elvis} et~al.,}{{Elvis}
  et~al.}{1994}]{Elvis_1994}
{Elvis} M.,  et~al., 1994, \mn@doi [\apjs] {10.1086/192093}, \href
  {https://ui.adsabs.harvard.edu/abs/1994ApJS...95....1E} {95, 1}

\bibitem[\protect\citeauthoryear{{Frank}, {King}  \& {Raine}}{{Frank}
  et~al.}{2002}]{Frank_2002}
{Frank} J.,  {King} A.,   {Raine} D.~J.,  2002, {Accretion Power in
  Astrophysics: Third Edition}.
Cambridge: Cambridge University Press

\bibitem[\protect\citeauthoryear{{Gierli{\'n}ski} \& {Done}}{{Gierli{\'n}ski}
  \& {Done}}{2004}]{Gierlinski_Done_2004}
{Gierli{\'n}ski} M.,  {Done} C.,  2004, \mn@doi [\mnras]
  {10.1111/j.1365-2966.2004.07687.x}, \href
  {https://ui.adsabs.harvard.edu/abs/2004MNRAS.349L...7G} {349, L7}

\bibitem[\protect\citeauthoryear{{Greene}}{{Greene}}{1959}]{Greene_1959}
{Greene} J.,  1959, \mn@doi [\apj] {10.1086/146759}, \href
  {https://ui.adsabs.harvard.edu/abs/1959ApJ...130..693G} {130, 693}

\bibitem[\protect\citeauthoryear{{Hubeny}}{{Hubeny}}{1990}]{Hubeny_1990}
{Hubeny} I.,  1990, \mn@doi [\apj] {10.1086/168501}, \href
  {https://ui.adsabs.harvard.edu/abs/1990ApJ...351..632H} {351, 632}

\bibitem[\protect\citeauthoryear{{Ichimaru}}{{Ichimaru}}{1977}]{Ichimaru_1977}
{Ichimaru} S.,  1977, \mn@doi [\apj] {10.1086/155314}, \href
  {https://ui.adsabs.harvard.edu/abs/1977ApJ...214..840I} {214, 840}

\bibitem[\protect\citeauthoryear{{Jin}, {Done}  \& {Ward}}{{Jin}
  et~al.}{2017a}]{Jin2017a}
{Jin} C.,  {Done} C.,   {Ward} M.,  2017a, \mn@doi [\mnras]
  {10.1093/mnras/stx718}, \href
  {https://ui.adsabs.harvard.edu/abs/2017MNRAS.468.3663J} {468, 3663}

\bibitem[\protect\citeauthoryear{{Jin}, {Done}, {Ward}  \& {Gardner}}{{Jin}
  et~al.}{2017b}]{Jin2017b}
{Jin} C.,  {Done} C.,  {Ward} M.,   {Gardner} E.,  2017b, \mn@doi [\mnras]
  {10.1093/mnras/stx1634}, \href
  {https://ui.adsabs.harvard.edu/abs/2017MNRAS.471..706J} {471, 706}

\bibitem[\protect\citeauthoryear{{Kaaret}, {Feng}  \& {Roberts}}{{Kaaret}
  et~al.}{2017}]{Kaaret_2017}
{Kaaret} P.,  {Feng} H.,   {Roberts} T.~P.,  2017, \mn@doi [\araa]
  {10.1146/annurev-astro-091916-055259}, \href
  {https://ui.adsabs.harvard.edu/abs/2017ARA&A..55..303K} {55, 303}

\bibitem[\protect\citeauthoryear{{Kato}, {Fukue}  \& {Mineshige}}{{Kato}
  et~al.}{2008}]{Kato_2008}
{Kato} S.,  {Fukue} J.,   {Mineshige} S.,  2008, {Black-Hole Accretion Disks
  --- Towards a New Paradigm ---}.
Kyoto University Press

\bibitem[\protect\citeauthoryear{{Katz}}{{Katz}}{1977}]{Katz_1977}
{Katz} J.~I.,  1977, \mn@doi [\apj] {10.1086/155355}, \href
  {https://ui.adsabs.harvard.edu/abs/1977ApJ...215..265K} {215, 265}

\bibitem[\protect\citeauthoryear{{King}, {Pringle}  \& {Livio}}{{King}
  et~al.}{2007}]{King_2007}
{King} A.~R.,  {Pringle} J.~E.,   {Livio} M.,  2007, \mn@doi [\mnras]
  {10.1111/j.1365-2966.2007.11556.x}, \href
  {https://ui.adsabs.harvard.edu/abs/2007MNRAS.376.1740K} {376, 1740}

\bibitem[\protect\citeauthoryear{{Klepnev} \& {Bisnovatyi-Kogan}}{{Klepnev} \&
  {Bisnovatyi-Kogan}}{2010}]{Klepnev_Bisnovatyi-Kogan_2010}
{Klepnev} A.~S.,  {Bisnovatyi-Kogan} G.~S.,  2010, \mn@doi [Astrophysics]
  {10.1007/s10511-010-9132-y}, \href
  {https://ui.adsabs.harvard.edu/abs/2010Ap.....53..409K} {53, 409}

\bibitem[\protect\citeauthoryear{{Li}, {Ho}, {Ricci}  \& {Trakhtenbrot}}{{Li}
  et~al.}{2024a}]{Li2024a}
{Li} R.,  {Ho} L.~C.,  {Ricci} C.,   {Trakhtenbrot} B.,  2024a, \mn@doi [\apj]
  {10.3847/1538-4357/ad77a5}, \href
  {https://ui.adsabs.harvard.edu/abs/2024ApJ...975...50L} {975, 50}

\bibitem[\protect\citeauthoryear{{Li}, {Ricci}, {Ho}, {Trakhtenbrot}, {Kara},
  {Masterson}  \& {Arcavi}}{{Li} et~al.}{2024b}]{Li2024b}
{Li} R.,  {Ricci} C.,  {Ho} L.~C.,  {Trakhtenbrot} B.,  {Kara} E.,  {Masterson}
  M.,   {Arcavi} I.,  2024b, \mn@doi [\apj] {10.3847/1538-4357/ad7aed}, \href
  {https://ui.adsabs.harvard.edu/abs/2024ApJ...975..140L} {975, 140}

\bibitem[\protect\citeauthoryear{{Liang} \& {Nolan}}{{Liang} \&
  {Nolan}}{1984}]{Liang_Nolan_1984}
{Liang} E.~P.,  {Nolan} P.~L.,  1984, \mn@doi [\ssr] {10.1007/BF00176834},
  \href {https://ui.adsabs.harvard.edu/abs/1984SSRv...38..353L} {38, 353}

\bibitem[\protect\citeauthoryear{{Lightman} \& {Shapiro}}{{Lightman} \&
  {Shapiro}}{1975}]{Lightman_Shapiro_1975}
{Lightman} A.~P.,  {Shapiro} S.~L.,  1975, \mn@doi [\apjl] {10.1086/181815},
  \href {https://ui.adsabs.harvard.edu/abs/1975ApJ...198L..73L} {198, L73}

\bibitem[\protect\citeauthoryear{{Liu}, {Liu}, {Wang}, {Cheng}  \&
  {Yuan}}{{Liu} et~al.}{2025}]{Liu2025}
{Liu} M.,  {Liu} B.~F.,  {Wang} Y.,  {Cheng} H.,   {Yuan} W.,  2025, \mn@doi
  [\mnras] {10.1093/mnras/staf459}, \href
  {https://ui.adsabs.harvard.edu/abs/2025MNRAS.tmp..440L} {}

\bibitem[\protect\citeauthoryear{{Magdziarz}, {Blaes}, {Zdziarski}, {Johnson}
  \& {Smith}}{{Magdziarz} et~al.}{1998}]{Magdziarz_1998}
{Magdziarz} P.,  {Blaes} O.~M.,  {Zdziarski} A.~A.,  {Johnson} W.~N.,   {Smith}
  D.~A.,  1998, \mn@doi [\mnras] {10.1046/j.1365-8711.1998.02015.x}, \href
  {https://ui.adsabs.harvard.edu/abs/1998MNRAS.301..179M} {301, 179}

\bibitem[\protect\citeauthoryear{{Malkan}}{{Malkan}}{1983}]{Malkan_1983}
{Malkan} M.~A.,  1983, \mn@doi [\apj] {10.1086/160981}, \href
  {https://ui.adsabs.harvard.edu/abs/1983ApJ...268..582M} {268, 582}

\bibitem[\protect\citeauthoryear{{Malkan} \& {Sargent}}{{Malkan} \&
  {Sargent}}{1982}]{Malkan_Sargent_1982}
{Malkan} M.~A.,  {Sargent} W.~L.~W.,  1982, \mn@doi [\apj] {10.1086/159701},
  \href {https://ui.adsabs.harvard.edu/abs/1982ApJ...254...22M} {254, 22}

\bibitem[\protect\citeauthoryear{{Manmoto}, {Mineshige}  \&
  {Kusunose}}{{Manmoto} et~al.}{1997}]{Manmoto_1997}
{Manmoto} T.,  {Mineshige} S.,   {Kusunose} M.,  1997, \mn@doi [\apj]
  {10.1086/304817}, \href
  {https://ui.adsabs.harvard.edu/abs/1997ApJ...489..791M} {489, 791}

\bibitem[\protect\citeauthoryear{{Mewe}, {Lemen}  \& {van den Oord}}{{Mewe}
  et~al.}{1986}]{Mewe_1986}
{Mewe} R.,  {Lemen} J.~R.,   {van den Oord} G.~H.~J.,  1986, \aaps, \href
  {https://ui.adsabs.harvard.edu/abs/1986A&AS...65..511M} {65, 511}

\bibitem[\protect\citeauthoryear{{Mihalas} \& {Mihalas}}{{Mihalas} \&
  {Mihalas}}{1984}]{Mihalas_1984}
{Mihalas} D.,  {Mihalas} B.~W.,  1984, {Foundations of radiation
  hydrodynamics}.
New York: Oxford University Press

\bibitem[\protect\citeauthoryear{{Narayan} \& {Yi}}{{Narayan} \&
  {Yi}}{1994}]{Narayan_Yi_1994}
{Narayan} R.,  {Yi} I.,  1994, \mn@doi [\apjl] {10.1086/187381}, \href
  {https://ui.adsabs.harvard.edu/abs/1994ApJ...428L..13N} {428, L13}

\bibitem[\protect\citeauthoryear{{Narayan} \& {Yi}}{{Narayan} \&
  {Yi}}{1995}]{Narayan_Yi_1995}
{Narayan} R.,  {Yi} I.,  1995, \mn@doi [\apj] {10.1086/176343}, \href
  {https://ui.adsabs.harvard.edu/abs/1995ApJ...452..710N} {452, 710}

\bibitem[\protect\citeauthoryear{{Novikov} \& {Thorne}}{{Novikov} \&
  {Thorne}}{1973}]{Novikov_Thorne_1973}
{Novikov} I.~D.,  {Thorne} K.~S.,  1973, in Black Holes (Les Astres Occlus). pp
  343--450

\bibitem[\protect\citeauthoryear{{Petrucci} et~al.,}{{Petrucci}
  et~al.}{2013}]{Petrucci_2013}
{Petrucci} P.~O.,  et~al., 2013, \mn@doi [\aap] {10.1051/0004-6361/201219956},
  \href {https://ui.adsabs.harvard.edu/abs/2013A&A...549A..73P} {549, A73}

\bibitem[\protect\citeauthoryear{{Petrucci}, {Ursini}, {De Rosa}, {Bianchi},
  {Cappi}, {Matt}, {Dadina}  \& {Malzac}}{{Petrucci}
  et~al.}{2018}]{Petrucci_2018}
{Petrucci} P.~O.,  {Ursini} F.,  {De Rosa} A.,  {Bianchi} S.,  {Cappi} M.,
  {Matt} G.,  {Dadina} M.,   {Malzac} J.,  2018, \mn@doi [\aap]
  {10.1051/0004-6361/201731580}, \href
  {https://ui.adsabs.harvard.edu/abs/2018A&A...611A..59P} {611, A59}

\bibitem[\protect\citeauthoryear{{Rees}}{{Rees}}{1988}]{Rees_1988}
{Rees} M.~J.,  1988, \mn@doi [\nat] {10.1038/333523a0}, \href
  {https://ui.adsabs.harvard.edu/abs/1988Natur.333..523R} {333, 523}

\bibitem[\protect\citeauthoryear{{Remillard} \& {McClintock}}{{Remillard} \&
  {McClintock}}{2006}]{Remillard_McClintock_2006}
{Remillard} R.~A.,  {McClintock} J.~E.,  2006, \mn@doi [\araa]
  {10.1146/annurev.astro.44.051905.092532}, \href
  {https://ui.adsabs.harvard.edu/abs/2006ARA&A..44...49R} {44, 49}

\bibitem[\protect\citeauthoryear{{Rybicki} \& {Lightman}}{{Rybicki} \&
  {Lightman}}{1979}]{Rybicki_Lightman_1979}
{Rybicki} G.~B.,  {Lightman} A.~P.,  1979, {Radiative processes in
  astrophysics}.
Wiley

\bibitem[\protect\citeauthoryear{{Sakimoto} \& {Coroniti}}{{Sakimoto} \&
  {Coroniti}}{1981}]{Sakimoto1981}
{Sakimoto} P.~J.,  {Coroniti} F.~V.,  1981, \mn@doi [\apj] {10.1086/159005},
  \href {https://ui.adsabs.harvard.edu/abs/1981ApJ...247...19S} {247, 19}

\bibitem[\protect\citeauthoryear{{Shakura} \& {Sunyaev}}{{Shakura} \&
  {Sunyaev}}{1973}]{Shakura_Sunyaev_1973}
{Shakura} N.~I.,  {Sunyaev} R.~A.,  1973, \aap, \href
  {https://ui.adsabs.harvard.edu/abs/1973A&A....24..337S} {24, 337}

\bibitem[\protect\citeauthoryear{{Shapiro}, {Lightman}  \& {Eardley}}{{Shapiro}
  et~al.}{1976}]{Shapiro_1976}
{Shapiro} S.~L.,  {Lightman} A.~P.,   {Eardley} D.~M.,  1976, \mn@doi [\apj]
  {10.1086/154162}, \href
  {https://ui.adsabs.harvard.edu/abs/1976ApJ...204..187S} {204, 187}

\bibitem[\protect\citeauthoryear{{Soria}, {Wu}  \& {Kunic}}{{Soria}
  et~al.}{2008}]{Soria_2008}
{Soria} R.,  {Wu} K.,   {Kunic} Z.,  2008, in {Carpano} S.,  {Ehle} M.,
  {Pietsch} W.,  eds, X-rays From Nearby Galaxies. pp 48--51 (\mn@eprint
  {arXiv} {0711.2448}), \mn@doi{10.48550/arXiv.0711.2448}

\bibitem[\protect\citeauthoryear{{Sunyaev} \& {Titarchuk}}{{Sunyaev} \&
  {Titarchuk}}{1980}]{Sunyaev_Titarchuk_1980}
{Sunyaev} R.~A.,  {Titarchuk} L.~G.,  1980, \aap, \href
  {https://ui.adsabs.harvard.edu/abs/1980A&A....86..121S} {86, 121}

\bibitem[\protect\citeauthoryear{{Thorne} \& {Price}}{{Thorne} \&
  {Price}}{1975}]{Thorne_Price_1975}
{Thorne} K.~S.,  {Price} R.~H.,  1975, \mn@doi [\apjl] {10.1086/181720}, \href
  {https://ui.adsabs.harvard.edu/abs/1975ApJ...195L.101T} {195, L101}

\bibitem[\protect\citeauthoryear{{Walton}, {Nardini}, {Fabian}, {Gallo}  \&
  {Reis}}{{Walton} et~al.}{2013}]{Walton_2013}
{Walton} D.~J.,  {Nardini} E.,  {Fabian} A.~C.,  {Gallo} L.~C.,   {Reis} R.~C.,
   2013, \mn@doi [\mnras] {10.1093/mnras/sts227}, \href
  {https://ui.adsabs.harvard.edu/abs/2013MNRAS.428.2901W} {428, 2901}

\bibitem[\protect\citeauthoryear{{Wandel} \& {Petrosian}}{{Wandel} \&
  {Petrosian}}{1988}]{Wandel_Petrosian_1988}
{Wandel} A.,  {Petrosian} V.,  1988, \mn@doi [\apjl] {10.1086/185166}, \href
  {https://ui.adsabs.harvard.edu/abs/1988ApJ...329L..11W} {329, L11}

\bibitem[\protect\citeauthoryear{{Wen}, {Jonker}, {Stone}, {Zabludoff}  \&
  {Psaltis}}{{Wen} et~al.}{2020}]{Wen_2020}
{Wen} S.,  {Jonker} P.~G.,  {Stone} N.~C.,  {Zabludoff} A.~I.,   {Psaltis} D.,
  2020, \mn@doi [\apj] {10.3847/1538-4357/ab9817}, \href
  {https://ui.adsabs.harvard.edu/abs/2020ApJ...897...80W} {897, 80}

\bibitem[\protect\citeauthoryear{{Zoghbi}, {Fabian}  \& {Gallo}}{{Zoghbi}
  et~al.}{2008}]{Zoghbi_2008}
{Zoghbi} A.,  {Fabian} A.~C.,   {Gallo} L.~C.,  2008, \mn@doi [\mnras]
  {10.1111/j.1365-2966.2008.14078.x}, \href
  {https://ui.adsabs.harvard.edu/abs/2008MNRAS.391.2003Z} {391, 2003}

\makeatother
\end{thebibliography}








\bsp	
\label{lastpage}
\end{document}